\documentclass[11pt,a4paper]{article}
\usepackage[latin1]{inputenc}
\usepackage{amsmath}
\usepackage{amsthm}
\usepackage{amsfonts}
\usepackage{amsfonts,amsthm,latexsym,amsmath,amssymb,amscd,epsfig,psfrag,enumerate}
\usepackage{graphics,graphicx, bezier, float, color, hyperref}
\usepackage{amssymb,url}
\usepackage{multienum}%
\usepackage[table]{xcolor}
\usepackage{multicol,multirow}
\usepackage{graphicx}
\usepackage{fancyvrb}
\usepackage{parskip}
\usepackage[toc,page]{appendix}

\usepackage{hyperref}  
\usepackage{float}

\usepackage{subcaption} 
\usepackage{pdfpages}
\usepackage[top=2.7cm, bottom=2.7cm, left=1.5cm, right=1.5cm]{geometry}
\usepackage{xcolor}
\usepackage{lipsum}
\usepackage{blkarray}

\usepackage[none]{hyphenat}[section]

\numberwithin{equation}{section}
\numberwithin{table}{section}
\numberwithin{figure}{section}
\usepackage{multirow}
\color{black}  
\usepackage{mathptmx}  
\title{
	A Stochastic Model for Illiquid Stock Prices and its
	Conclusion about Correlation Measurement}
\author{Erina Nanyonga$^{\text{*}} $, Juma Kasozi$^{\text{*}}$, Fred Mayambala$^{\text{*}}$  Hassan W. Kayondo$^{\text{*}}$ and Matt Davison$^{\text{**}} $\\
	$^{\text{*}}$\textit{Department of Mathematics, Makerere University, Kampala, Uganda},\\$^{ \text{**}}$	\textit{Department of Statistical-Actuarial Sciences and Mathematics,
		Western University Canada}\\
	Corresponding author's email: \textit{erinananyonga9@gmail.com}}
\date{}
\begin{document}
	
	\maketitle
	\begin{abstract}
		\noindent
		\textbf{Abstract}
		
		\noindent	
		This study explores the behavioral dynamics of illiquid stock prices in a listed stock market. Illiquidity, characterized by wide bid and ask spreads affects price formation by decoupling prices from standard risk and return relationships and increasing sensitivity to market sentiment. We model the prices at
		the Uganda Securities Exchange (USE) which is illiquid in  that the prices remain constant much of the time thus complicating  price modelling. We circumvent this challenge by combining the  Markov model (MM) with two models;  the exponential Ornstein Uhlenbeck model (XOU) and geometric Brownian motion (gBm). In the combined models, the MM was used to capture the constant prices  in the stock prices while the XOU and gBm   captured the stochastic price dynamics. We modelled stock prices using the combined models, as well as XOU and gBm alone. We found that USE stocks appeared to have low correlation with one another. Using theoretical analysis, simulation study and empirical analysis, we conclude that this apparent low correlation is due to illiquidity. In particular data simulated from combined MM-gBm, in which the gBm portion were highly correlated resulted in a low measured correlation when the Markov chain had a higher transition from zero state to zero state.
		
		\vspace{1em}
		\noindent\text{Key words:} Illiquid market, Markov model, exponential Ornstein Unlenbeck , Uganda Securities Exchange, geometric Brownian motion, Kolmogorov-Smirnov test, rolling correlation
	\end{abstract}	
	
	\setlength{\parindent}{20pt}  
	\setlength{\parskip}{0pt}     
	
	\vspace{0.2cm}
	\section{Introduction}

Quantitative Finance often takes as a foundational assumption that markets are deep and liquid - which means you can always see a price, and always trade your desired quantity of securities at that price.    In some markets, for instance the USD/EUR foreign exchange market,  the front contract of crude oil, or the biggest stocks trading on the NYSE,  this assumption is  reasonable but not all securities are this liquid.  For instance,  on the Uganda Securities Exchange (USE)
often stock prices do not change from one day to the next.
Such stock prices are hard to model with stochastic differential equations (SDEs),  which suggest that price moves are drawn from a continuous return distribution from which it is vanishingly likely to draw a return of exactly zero.

In this paper, we present and analyze  innovative models for the random fluctuation of stock prices. Our models are based on a  Markov model (MM) which decides if stock prices move at all, coupled with an underlying SDE-based model which measures how much the prices move.   We fit the models to several Ugandan Stocks and discuss the results.

In an illiquid market, the market process is not completely observable, because no trades are completed on some days. This allows changes in market fundamentals and market sentiment alike to remain uncaptured by price data.    We model whether trades occur or not with a Markov chain that measures whether transactions are completed or not \cite{ccanakouglu2011portfolio}.  
A Markov chain is based on the assumption that the current state alone is sufficient to predict the next state. In a similar way, the random walk hypothesis \cite{fama1965behavior} also makes the same Markov assumption.

According to \cite{saad1998comparative} 
the stock market is governed by a mixture of both deterministic and random factors.  Stock prices are influenced by economic factors (such as inflation, currency exchange rates, policy adjustments),  disease outbreaks (COVID 19),  international relations (wars and diplomatic incidents), and others.  The interplay of these complicated trends is hard to understand. Stock prices are also influenced by news idiosyncratic to the given company and by the dynamics of trading on the stock market itself (a big purchase of stocks drives the price up, a big sale depresses the price).  The impact of all of these factors is best modelled by a random component.  
Characterizing the random processes driving a given stock price is very important because it allows the construction of stock portfolios,  the hedging of stock options, and the accomplishment of many risk management tasks.

Brownian motion is used to provide randomness in
stock markets, foreign exchange markets, commodity
markets and bonds.
Brownian motion goes negative but asset prices cannot be negative in the real world scenario. 
Many price models are SDEs including geometric Brownian motion (gBm), Ornstein Uhlenbeck model, Cox Ingersoll Ross model, geometric mean reverting model and others. Other stochastic models include time series models such as auto regressive
integrated moving average, generalized autoregressive
conditional heteroskedasticity and others.

A market in which it is difficult to trade assets at desired quantities is termed an ``illiquid'' market.  A market may be illiquid because of a temporary absence of  buyers or sellers in response to various market events, or it may be permanently illiquid because the product being traded is just simply not of sufficient interest to investors.  Illiquid markets, with their low levels of trading activity,  are often characterized by wider spreads between bid and offer (ask) prices, particularly for large trades.  It should be noted that liquidity is partly a scale dependent phenomenon, it may be perfectly easy to buy or sell small orders of something with low transaction costs, but  difficult to fill large orders.
Although USE has some large
listed companies  for instance Standard Bank Uganda with a current market capitalization of \$729.22M, the market is quite illiquid. The data indicate that nearly all USE stocks are illiquid when
compared with large stocks in developed markets. For example, even for the most liquid stock, there exists a number of non-trading days for it and a number of consecutive days with a constant stock price, this is observed in Figure~\ref{fig1}.

When many of people come to the
exchange to trade, that's a process driven by Brownian motion where prices will be randomly moving up and down. When they don't come,
it's something else driving the process since prices become constant and stop moving up and down. The market becomes illiquid when fewer people are entering it, or no much
activity as it is at USE much of the time, or when no body wants to buy the available stocks. USE is not the only African stock exchange that is illiquid, there are other illiquid exchanges  such as Lusaka Stock Exchange, Nairobi Securities Exchange, Dar es Salaam Stock Exchange. However, every illiquid market is illiquid in its own way and in this research, we will study the illiquidity of stocks at USE.

\cite{kasozi2023prediction} made
significant strides   in understanding the stock dynamics at USE,  where they used the exponential Ornstein Uhlenbeck  model (XOU) to predict stock prices, but several key gaps remained. One important gap is the lack of studies examining the process driving the  constant prices much of the time at the exchange. The predicted  prices before the COVID 19 outbreak in their work never had constant values at all yet the actual ones exhibited them. This was even worse during the outbreak of COVID 19, their predicted values were all moving up and down yet the actual ones were constant much of the time. This implies that their predictions were not so close to the USE market scenario with volatile and constant prices much of the time.
Understanding the modelling of constant values much of the time in the USE market stocks remained 
a challenge. Filling this gap will contribute to the development of more accurate predictive models at USE.

In every market, there are some securities that do not trade very much, and those that  trade more frequently. Being an illiquid market, the stocks at USE have varying non-trading days each year, and can be so many for some stocks and very few for others hence, trading days for each security at USE differ. Understanding the price dynamics of the stocks at USE  can yield higher returns for investors who allocate their capital in such an illiquid market. 
To capture the stochastic stock dynamics at USE, SDEs that is; gBm and XOU were used. The XOU is a log normal process hence, good approximations for the volatility distribution. 
\cite{perello2008option} 
points it out that the 
XOU, is able to describe simultaneously the observed long-range memory
in volatility and the short-range one in leverage which is more adherent to economic reality. It is able to provide a
consistent stationary distribution for the volatility with data, that is; it reliably yields results that align well with the observed data over time and lastly, the XOU
reproduces the realized volatility fairly well hence,  being able to capture stock price fluctuations in the future.  The gBm is the simplest SDE with two parameters while the XOU is a mean reverting model with three parameters. 
The identified gap was  filled 
by the MM since it allows an action to be repeated several times, the up and down movement was captured by the SDEs. We modelled the stock prices at USE using the combined models as well as the XOU and gBm alone.

We computed the rolling window correlation of pairs of stock prices at USE. The rolling window correlation is important in financial markets due to changes in relationships between two assets over time which may be caused by varying market conditions and  economic factors.
The computation of rolling window correlation enables one to observe the changing relationship that exists between two variables over time, putting in mind that the relationships between assets especially in mature markets is not constant. Strong correlation of assets in a portfolio can be used to represent price movements of each other.

A strong correlation among specific groups of assets in a financial market would suggest that the market is influenced by shared economic factors among the assets.
The correlation between assets in a portfolio is crucial for diversification, particularly when aiming to minimize exposure to sector or industry specific breakdowns \cite{rosen2006correlation}.
Research demonstrates that the price trends of some stocks have been predictable using information about the price trends of other stocks. 

In order to find out the effect of the constant stock prices much of the time on correlation of Ugandan stocks, we carried out a theoretical and simulation study using the combined model of MM and gBm.  We computed the correlation on simulated data when the number of repetitions in the prices were high and low and measured the correlation of the simulated prices based on steady state probabilities and two correlated gBm.
\section{Methods and Materials}
\subsection{Data Used}

Data for the listed stocks that we considered at USE was freely  downloaded from the Wall Street Journal (WSJ) website; \url{https://www.wsj.com/market-data/quotes/UG/XUGA/SBU/historical-prices} 
for Stanbic Bank Uganda (SBU), 
\url{https://www.wsj.com/market-data/quotes/UG/XUGA/BOBU/historical-prices} for bank of Baroda Uganda (BOBU) historical prices,
and
\url{https://www.wsj.com/market-data/quotes/UG/XUGA/DFCU/historical-prices} for 	Development Finance Company of Uganda Ltd (DFCU) historical data.
Data for the USE  All Share Index (ALSIUG), the main stock market index in Uganda, was downloaded from investing.com on the following url; \url{https://www.investing.com/indices/uganda-all-share-historical-data}. For the stocks that are cross listed on the Nairobi Securities Exchange (NSE) and USE; Equity Group Holdings (EQTY), \url{https://www.wsj.com/market-data/quotes/KE/XNAI/EQTY/historical-prices} and Kenya Commercial Bank (KCB); \url{https://www.wsj.com/market-data/quotes/KE/XNAI/KCB/historical-prices}

\subsection{The Models and Detailed Analysis} 

	\subsubsection{ The exponential Ornstein-Uhlenbeck model (XOU)}
The price $\{S(t), t\in\mathbb{R}^{+}\}$  is said to follow the XOU process if it satisfies the following SDE;

\begin{equation}\label{eqn1}dS(t)=\gamma(\phi -\text{log}\hspace{0.09cm}S(t))S(t)\,dt+\sigma S(t)\,dW(t),\end{equation}

where $\sigma > 0$ is the volatility term, $\phi$ is the process long-term expected value,  $\gamma > 0$ is the speed (reversion of $ S(t)$ towards $\phi$), all assumed to be constant and $dW(t)$ is an increment during the interval $(t,t+dt)$ of a standard Brownian motion. 
This model appears in many studies such as; \cite{schwartz1997stochastic},  \cite{mejia2018calibration}, \cite{kasozi2023prediction}.
Working with logarithms of the closing stock prices, the solution to the XOU is given by \cite{kasozi2023prediction};

\begin{equation}\label{eqn2}
	\text{log}\hspace{0.1cm}S(t)=\text{log}\hspace{0.1cm}S(\tau)\,e^{-\gamma(t-\tau)}+\left(\phi -\frac{\sigma ^{2}}{2\gamma}\right)(1-e^{-\gamma(t-\tau)})+\,\sigma \int_{\tau}^{t}e^{-\gamma(t-\bar{t})}\,dW(\bar{t}).\end{equation}

The discretized form of Equation~\eqref{eqn2} is given by;

\begin{equation}\label{eqn3}
	\text{log}\hspace{0.1cm}S(t)=\text{log}\hspace{0.1cm}S(\tau)\,e^{-\gamma(t-\tau)}+\left(\phi -\frac{\sigma ^{2}}{2\gamma}\right)(1-e^{-\gamma(t-\tau)})+\sigma\sqrt{\frac{1}{2\gamma}\left[1-e^{-2\gamma(t-\tau)}\right]
	} \hspace{0.1cm}\xi (t),
\end{equation}  

$\text{with an error term,}\hspace{0.1cm}\xi (t)\sim \mathcal{N}(0,1)$
and

$$\text{log}\hspace{0.1cm}S(t)\sim \mathcal{N}\bigg(\text{log}\hspace{0.1cm}\hspace{0.1cm}S(\tau)\,e^{-\gamma(t-\tau)}+\left(\phi -\frac{\sigma ^{2}}{2\gamma}\right)(1-e^{-\gamma(t-\tau)}),\frac{\sigma ^{2}}{2\gamma}\left[1-e^{-2\gamma(t-\tau)}\right]\bigg).$$
The equation of the conditional probability density of $\text{log}\hspace{0.1cm}S(t)$ under the XOU is given by

\begin{equation}\label{eqn4}
	f(\text{ln}\hspace{0.09cm}S(t)|\text{ln}\hspace{0.09cm}S(\tau);\phi, \gamma,\sigma)=\frac{1}{\sqrt{2\pi \hat{\hat{\sigma}} ^{2}}}\,\text{exp}\left(-\frac{\left(\text{ln}\hspace{0.09cm}S(t)-\left(\text{ln}\hspace{0.09cm}S(\tau)\,e^{-\gamma(t-\tau)}+\hat{\hat{\phi}}(1-e^{-k(t-\tau)})\right)\right)^{2}}{2\hat{\hat{\sigma}}^{2}}\right), 
\end{equation}

where $\hat{\hat{\phi}}=\left(\phi -\frac{\sigma ^{2}}{2\gamma}\right)\hspace{0.2cm}\text{and}\hspace{0.2cm}\hat{\hat{\sigma}}^2=\hspace{0.2cm}\frac{\sigma ^{2}}{2\gamma}\left[1-e^{-2\gamma(t-\tau)}\right]\hspace{0.1cm}\text{for}\hspace{0.2cm} t>\tau.$
Considering the time step as $\Delta t$, 	and

$$Y_{2}=\sum_{i=1}^{n}Y_{i},~~ Y_{1}=\sum_{i=1}^{n}Y_{i-1},~~ Y_{1,1}=\sum_{i=1}^{n}Y^{2}_{i-1},~~
Y_{1,2}=\sum_{i=1}^{n}Y_{i-1}Y_{i},~~
Y_{2,2}=\sum_{i=1}^{n}Y^{2}_{i},$$

where $Y_i$ is  the natural log of the prices at  time  $i$ and $Y_{i-1}$ the natural logarithm of the prices at time $i-1$ the parameters of the XOU using the maximum likelihood estimator are;

\begin{eqnarray}\label{eqn5}
	\hat{\hat{\phi}}&=&\frac{Y_{2}Y_{1,1}-Y_{1}Y_{1,2}}{n(Y_{1,1}-Y_{1,2})-(Y^{2}_{1}-Y_{2}Y_{1})},\end{eqnarray}

\begin{equation}	\label{eqn6}
	\gamma= -\frac{1}{\triangle t}\hspace{0.09cm}\text{ln}\hspace{0.09cm}\left(\frac{Y_{1,2}-\hat{\hat{\phi}}Y_{1}-\hat{\hat{\phi}}Y_{2}+n\hat{\hat{\phi}}^{2}}{Y_{1,1}-2\hat{\hat{\phi}}Y_{1}+n\hat{\hat{\phi}}^{2}}\right),
\end{equation}

\begin{equation}\label{eqn7}
	\hat{\hat{\sigma}}^{2}
	=\frac{1}{n}[Y_{2,2}-2e^{-\gamma\triangle t}Y_{1,2}+e^{-2\gamma\triangle t}Y_{1,1}-2\hat{\hat{\phi}}(1-e^{-\gamma\triangle t})(Y_{2}-e^{-\gamma\triangle t}Y_{1})+n\,\hat{\hat{\phi}}^{2}(1-e^{-\gamma\triangle t})^{2}],
\end{equation}

\begin{eqnarray}\label{eqn8} \hspace{0.5cm}	\sigma ^{2}&=&\hat{\hat{\sigma}}^{2}\frac{2\gamma}{1-(e^{-\gamma\triangle t})^{2}},\\\nonumber
	\text{and from}\hspace{0.5cm}
	\hat{\hat{\phi}} &=&\phi
	-\frac{\sigma ^{2}}{2\gamma},\hspace{0.5cm}\text{we have}\\\label{eqn9}  \phi &=&\hat{\hat{\phi}}
	+\frac{\sigma ^{2}}{2\gamma}.\end{eqnarray}
The constant values at USE were captured by the MM.
\subsubsection{The Markov model (MM)} 
A Markov chain is a (Markov) model
representing the probabilities of sequences of random variables,
states, each of which can take on values from some set. 
Consider a sequence of state variables $q_1,q_2,q_3,\ldots, q_i$. A Markov
model incorporates the Markov assumption on the probabilities of this sequence: that
when predicting the future, the past does not matter, only the present does. That is; 

$$\mathbb{P}(q_{i}=a\mid q_1,\ldots ,q_{i-1})=\mathbb{P}(q_{i}=a\mid q_{i-1})$$

A Markov chain is characterized by the following components;
\begin{itemize}
	\item Number of finite states in the model, $N$. 
	\item A transition probability matrix $A,$ each $a_{ij}$ representing the probability of moving from state $i$ to state $j,$ such that  for all $i,$ 
	$$\sum_{j=1}^{n} a_{ij}=1$$
	\item An initial probability distribution over states. $\pi_i$ is the
	probability that the Markov chain will start in state $i.$
\end{itemize}

\subsubsection{Combination of MM and XOU }
We consider  a 2 state MM model for the stock prices; 
that is, a price change taken as state 1 and, a constant price taken as state 0; State 1 is controlled by  Equation \eqref{eqn1}. 
A combination of the MM and the XOU models (MM-XOU) was described by; 

$$
dS(t) =
\begin{cases}
	0 & \text{if} ~~~~~S(t+1) = S(t), \\
	\gamma(\phi -\text{log}\,S(t))dt+\sigma S(t) dW(t)& \text{if} ~~~~~S(t+1) \neq S(t).
\end{cases}
$$

When the price is constant,  the movement described by  0, while if there is a movement in the stock price, the XOU describes it. Hence
a random 2 state series was generated using the MM for a change in price movement and no change at all;

$$
\text{States for  prices} =
\begin{cases}
	0 & \text{if} ~~~~~S(t+1)= S(t), \\
	1& \text{if} ~~~~~S(t+1)\neq S(t).
\end{cases}
$$

The stock prices can move from state 0 to 0 with a probability $p$, from 0 to 1 with probability $1- p$, from 1 to 1 with probability $q$ and from 1 to 0 with a probability of $1- q$ where
the state probabilities $p $ and $q$ are;

\begin{equation}\label{eqn10}
	p=\frac{\text{counts}[0]}{ (\text{counts}[0] + \text{counts}[1])},\end{equation}
\begin{equation}\label{eqn11}
	q=\frac{\text{counts}[3]}{ (\text{counts}[2] +\text{counts}[3])},
\end{equation}
where transitions from states are represented by; { ``{0$\rightarrow$0}": 0, ``{0$\rightarrow$1}": 1, ``{1$\rightarrow$0}": 2, ``{1$\rightarrow$1}": 3}.
Hence the transition probabilities matrix is given by;

$$
A = \begin{bmatrix}
	p & 1-p \\
	1-q & q
\end{bmatrix}
$$

where
\begin{itemize}\item $p = \mathbb{P}(X_{n+1} = 0 \mid X_n = 0)$ is the probability of \text{staying} in state 0 and 
	\item $q = \mathbb{P}(X_{n+1} = 1 \mid X_n = 1)$ is the probability of \text{staying} in state 1.
\end{itemize}

\subsubsection{ Geometric Brownian motion (gBm)}
A stock price $S(t)$ follows gBm if it satisfies Equation~\eqref{eqn12}.

\begin{equation}\label{eqn12}
	dS=\mu S(t)dt+\sigma S(t)dW(t),
\end{equation} 

where $\mu$ is the drift term and $\sigma$ the volatility term.
The parameters of the gBm  are obtained as below;

\begin{eqnarray}\label{eqn13}
	\text{Estimated drift,}~~\hat{\mu} &=& \frac{1}{m}\sum_{i=1}^{m}R(i),\\\label{eqn14}
	\text{Estimated volatility,}~~ \hat{\sigma}  &=& \sqrt{\frac{1}{m-1} \sum_{i=1}^{m}(R(i)-\hat{\mu})^2},\end{eqnarray}
using logarithms of prices	where    $ R(i)=\text{ln}\left(\frac{S(i)}{S(i-1)}\right)$ is the return at time $i$.  Studies that utilized gBm for stock price prediction include; \cite{samuelson1965}, \cite{abidin2012review} and \cite{liden2018stock} and others. Similar to the XOU, gBm also does not allow negative prices. According to  \cite{abidin2012review}, gBm  works best for  short-term
investment with easier calculations despite its weakness of considering considering constant volatility.

\subsubsection{Combination of MM and gBm}
We consider the combination of MM and gBm as MM-gBm where the change in the stock prices follows a gBm as described below;

$$
dS(t) =
\begin{cases}
	0 & \text{if} ~~~~~S(t+1) = S(t), \\
	\mu S(t)dt+\sigma S(t)dW(t)& \text{if} ~~~~~S(t+1) \neq S(t).
\end{cases}
$$

When the price is constant, the movement described by  0, while if there is a movement in the stock price, the gBm describes it. A random 2 state series was generated using the MM for a change in price movement and no change at all;

$$
\text{States for  prices} =
\begin{cases}
	0 & \text{if} ~~~~~S(t+1) = S(t), \\
	1& \text{if} ~~~~~S(t+1) \neq S(t).
\end{cases}
$$

The stock prices can move from state 0 to 0 with a probability $p$, from 0 to 1 with probability $1- p$, from 1 to 1 with probability $q$ and from 1 to 0 with a probability of $1- q$ where
the state probabilities $p $ and $q$  are described by Equations \eqref{eqn10} and \eqref{eqn11}.

\subsection{Performance Measure}
We compared the forecasting accuracy of the four models using the mean absolute percentage error (MAPE) in each time period, for each selected security at USE.

\begin{equation}\label{eqn15}
\text{MAPE} = \frac{1}{n} \sum_{i= 1}^{n}\frac{|S_i -\hat{S_i}|}{S_i}\times 100 \%
\end{equation}

where $S_i$ represents the actual closing price, $\hat{S_i}$ represents the forecasted closing price and $i=1,2,\ldots, n.$

\subsection{Programming Software} We used Sagemath 9.5 in all our simulations. The computer used had Ubuntu 24.04.2 LTS operating system, Intel Core i5-6300U processor, 8.0GiB memory and 256.1GB disk capacity.

\subsection{ Correlation Measurement}We utilized the Pearson correltion coefficient to compute correlation of stocks.
For two time series $X = \{x_1, x_2, \ldots, x_n\}$ and $Y = \{y_1, y_2, \ldots, y_n\}$, 
Pearson correlation coefficient is given by; 
\begin{equation}
\rho= \frac {\displaystyle \sum_{i=1}^n (x_i - \bar{x})(y_i - \bar{y})}{\displaystyle \sqrt{\sum_{i=1}^n (x_i - \bar{x})^2} \sqrt{\sum_{i=1}^n (y_i - \bar{y})^2}}
\end{equation}
where 
$	x_i, y_i$
are the data points with mean
$
\bar{x}, \bar{y}
$ respectively for
$ n $   number of observations.
The rolling window correlation of window size, $k$ at position $ i $ is defined as:
\begin{equation}
\rho_k = \frac{\displaystyle
	\sum_{j=i}^{i+k-1} \left(x_j - \bar{x}_{i}\right) \left(y_j - \bar{y}_{i}\right)	}{\displaystyle
	\sqrt{ \sum_{j=i}^{i+k-1} \bigl(x_j - \bar{x}_{i}\bigr)^2
	}
	\sqrt{\sum_{j=i}^{i+k-1} \left(y_j - \bar{y}_{i}\right)^2
	}
}
\end{equation}
where 
$
\bar{x}_{i} = \frac{1}{k} \sum_{j=i}^{i+k-1} x_j, \quad
\bar{y}_{i} = \frac{1}{k} \sum_{j=i}^{i+k-1} y_j
,$  the starting index of the rolling window, $i = 1, 2, \ldots, n-k+1$ and $j$ is the index variable used within each window.

\section{Results}

Model calibration is  as important as the model itself. It involves determining the parameter values that allow the model to replicate market prices with the highest possible accuracy. Both the precision and efficiency of calibration are essential, as practitioners rely on the calibrated parameters to price numerous complex derivative contracts and create high-frequency trading strategies \cite{cui2017full}.  Once the model's analytical solution is known, the calibration problem is solved by determining the parameter values that yield the ``best fit" between the model and the data used for calibration \cite{fatone2024calibration}. The more accurate the calibration, the greater the predictive validity of the model, making it a more valuable tool for risk management and portfolio optimization. 

\begin{figure}[h]
	\centering
	\subfloat[ BOBUprices from $7/5/2023-7/5/2024.$.]{%
		\resizebox*{6cm}{!}{\includegraphics{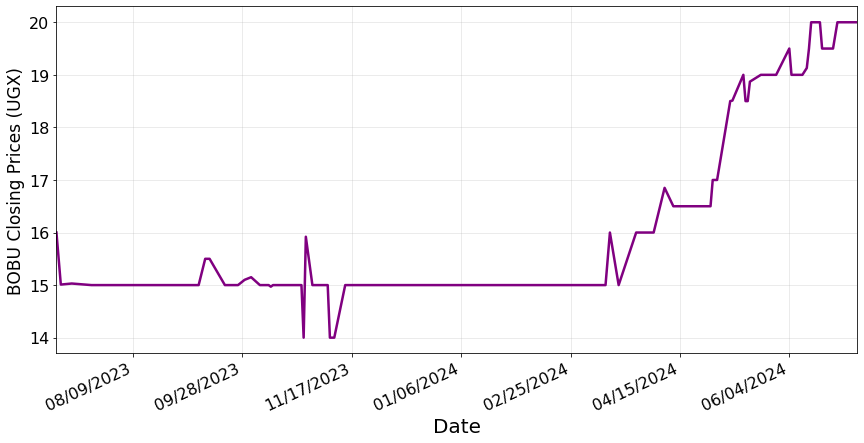}}}\hspace{5pt}
	\hspace{1cm}
	\subfloat[SBU prices from $1/3/2023-12/28/2023.$.]{%
		\resizebox*{6cm}{!}{\includegraphics{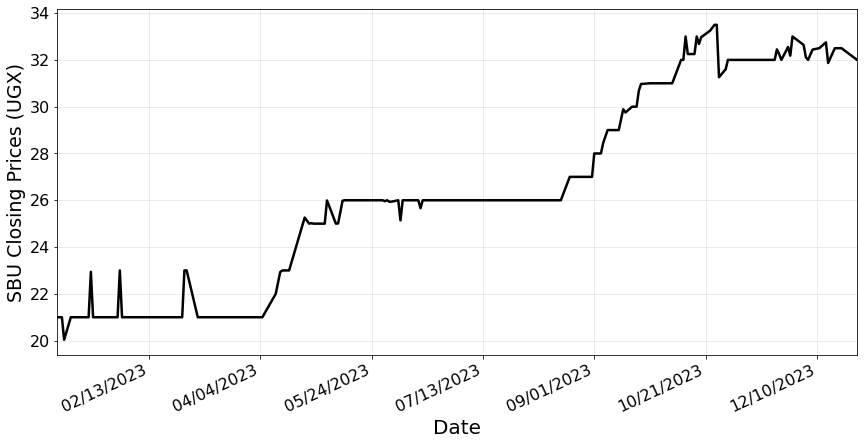}}}\hspace{5pt}
	
	\subfloat[DFCU prices from $1/17/2023-12/23/2024.$]{%
		\resizebox*{6cm}{!}{\includegraphics{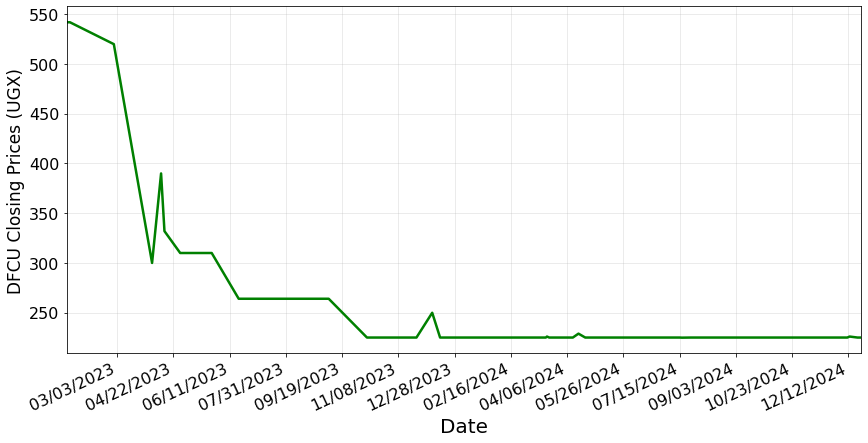}}}\hspace{5pt}
	\hspace{1cm}
	\subfloat[ALSIUG from $1/2/2024-12/31/2024. $.]{%
		\resizebox*{6cm}{!}{\includegraphics{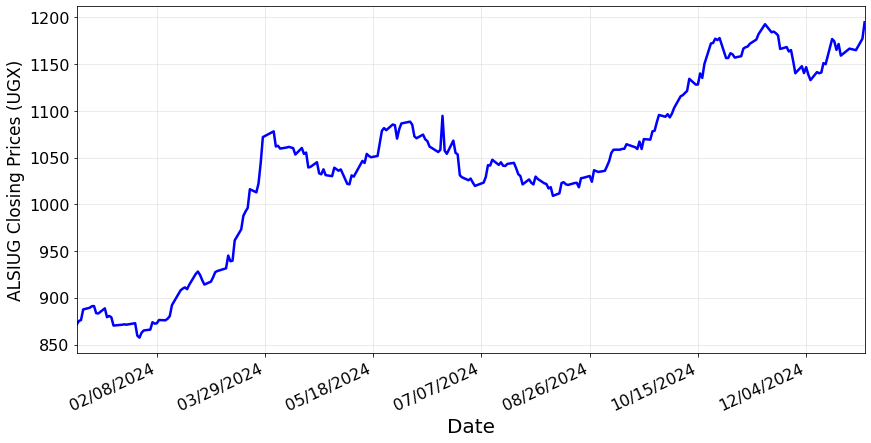}}}\hspace{5pt}
	\caption{Actual closing prices of some of the securities at USE.}\label{fig1}
\end{figure}

We modelled closing prices of securities  at USE that is; SBU, BOBU, DFCU and ALSIUG; 
using the gBm, MM-gBm, XOU and the MM-XOU.  In Figure~\ref{fig1}, the closing prices of the stocks  are constant much of the time and always move up and down for ALSIUG, the main stock market index in Uganda which tracks the performance of most of the stocks listed at the exchange. 

\begin{figure}[h]
	\centering
	\subfloat[SBU.]{%
		\resizebox*{4cm}{!}{\includegraphics{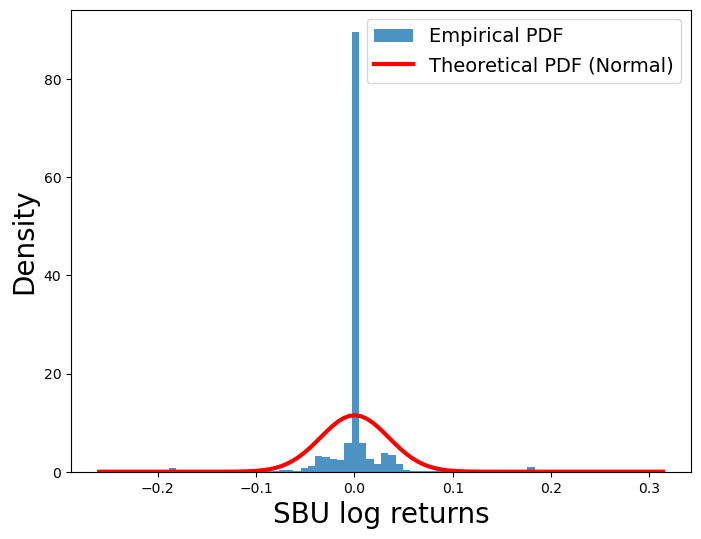}}}\hspace{5pt}
	\hspace{1cm}
	\subfloat[BOBU.]{%
		\resizebox*{4cm}{!}{\includegraphics{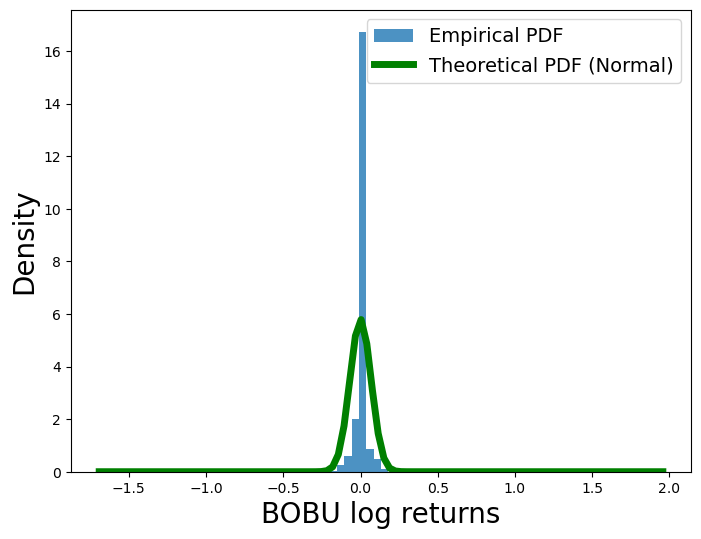}}}\hspace{5pt}
	
	\subfloat[DFCU.]{%
		\resizebox*{4cm}{!}{\includegraphics{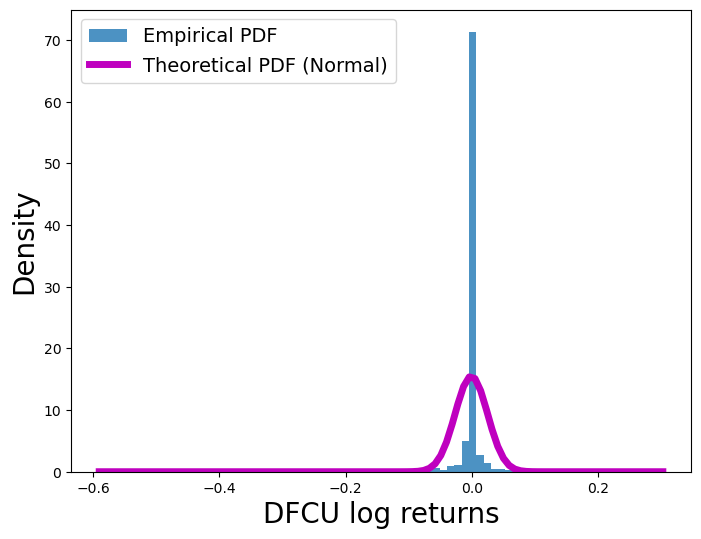}}}\hspace{5pt}
	\hspace{1cm}
	\subfloat[ALSIUG.]{%
		\resizebox*{4cm}{!}{\includegraphics{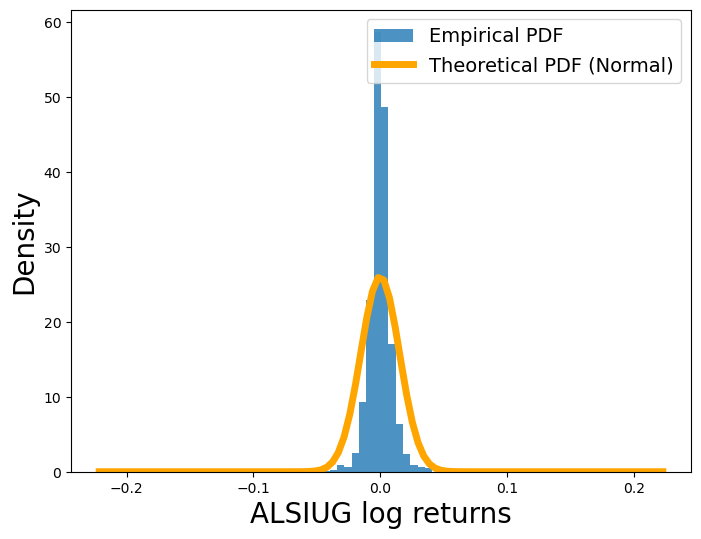}}}\hspace{5pt}
	\caption{Empirical versus theoretical PDF of logarithmic returns of securities at USE.}\label{fig2}
\end{figure}

Small fluctuations dominate the  returns of some of the securities at USE as illustraded in Figure~\ref{fig2}, where SBU data from $ 6/28/2011 - 12/6/2024 $, BOBU data from $ 7/18/2011 - 12/10/2024 $, DFCU data from $ 6/28/2011 - 12/11/2024 $, ALSIUG data from $ 7/17/2017 - 1/16/2025 $ was utilized to plot the empirical PDF of their logarithmic returns against the theoretical normal PDF.

Figure~\ref{fig2} provides the probabilities of a range of possible returns hence, we can forecast  the likelihood of future returns of the securities which helps in making investment decisions. 

We modelled the stock prices of SBU before, during and after COVID 19 using a two, three, four and five parameter model that is; gBm,  XOU, MM-gBm and MM-XOU respectively.
Different time periods were used to calibrate the models in each case.  
The parameters were obtained using Equations \eqref{eqn6}, \eqref{eqn8} and \eqref{eqn9} for the XOU. 
The gBm parameters were obtained using Equations \eqref{eqn13} and \eqref{eqn14}. The calibration of the XOU and gBm was done using the original data of the closing daily prices of the securities at USE (uncleaned data). 

Repetitions were removed from the data used for calibration of MM-XOU and MM-gBm, we considered this as cleaned data. There were 2 possible random states for the prices to undertake; 0 and 1, the state probabilities $p $ and $q$  were obtained using Equations \eqref{eqn10} and \eqref{eqn11}. We modelled closing prices of ALSIUG and a few of the stocks. In  all the tables with MAPE values results, we let Sn represent simulation, with Sn\_1, Sn\_2 and Sn\_3 representing the first, second and third simulations respectively.

\subsection{Modelling SBU closing prices before COVID 19}
Closing prices of SBU before COVID 19 were modelled for two different time periods for 100, 50 and 30 days in each case. The first period started from  $ 6/29/2017 - 12/21/2017 $ while the second one started from $ 1/9/2018 - 7/25/2018 $.  Data for calibration of the models in the first time period started from $ 7/27/2016 - 6/28/2017 $ (a total of 200 days), and for the second time period, calibration data started from $ 1/16/2017 - 1/5/2018 $ for all the four models. All these 200 days were utilized to calibrate gBm and XOU. In the first and second  time periods, 73 days remained after cleaning the data and were used to calibrate MM-XOU and MM-gBm. The  MAPE values from three different simulations of each model at each time are summarized in Table~\ref{tab1}.
The parameters obtained were the following;
$$\begin{array}{lrrrrrr}
	\text{MM-gBm}&:&\mu=-0.0005,&\sigma=0.0334,&p=0.7143,&q=0.5&\\
	\text{gBm}&:&\mu=-0.0002,&\sigma=0.02&&&\\
	\text{MM-XOU}&:&\gamma=0.0035,&\sigma=0.033,&\phi=3.5042,&p=0.7143,&q=0.5\\
	\text{XOU}&:&\gamma=0.0035,&\sigma=0.0199,&\phi=3.3426
\end{array}
$$
\begin{table}[h]\scriptsize \centering
	\caption{MAPE  at different time periods while modelling SBU stock prices before COVID 19 in the first time period that was selected ($ 6/29/2017 - 12/21/2017 $).}	
	{	\begin{tabular}{ccccccc}
			\hline 
			\multicolumn{3}{c}{}                                                                                                                               & \multicolumn{4}{c}{\text{MAPE (\%) from different simulations}}\\ \hline
			\text{Date}        &\text{Total  days} & \text{Sn}    & \text{MM- gBm}      & \text{gBm}          & \text{MM- XOU}      & \text{XOU}   \\
			\hline
			\multirow{3}{*}{\begin{tabular}[c]{@{}c@{}}6/29/2017\\-\\12/21/2017\end{tabular}}  & \multirow{3}{*}{100}                                   & Sn\_1 &  1.0793                                            & 1.0077 & 0.3958                                            & 0.5272 \\
			&                                                       & Sn\_2 & 1.0546                                            & 0.7007 & 0.4789                                            & 0.4764 \\
			&                                                       & Sn\_3 & 1.1512                                            & 1.3059 & 0.575                                             & 0.5504 \\
			\hline
			\multirow{3}{*}{\begin{tabular}[c]{@{}c@{}}6/29/2017\\ -\\ 9/20/2017\end{tabular}}   & \multirow{3}{*}{50}                                   & Sn\_1 & 0.2691                                            & 0.8487 & 0.4089                                            & 0.4872 \\
			&                                                       & Sn\_2 & 0.347                                             & 0.8125 & 0.3598                                            & 0.375  \\
			&                                                       & Sn\_3 & 0.3481                                            & 0.7948 & 0.4589                                            & 0.4913 \\
			\hline
			\multirow{3}{*}{\begin{tabular}[c]{@{}c@{}}6/29/2017\\ -\\ 8/16/2017\end{tabular}}   & \multirow{3}{*}{30}                                   & Sn\_1 & 0.2813                                            & 0.4936 & 0.4411                                            & 0.4327 \\
			&                                                       & Sn\_2 & 0.253                                             & 0.4827 & 0.4386                                            & 0.4153 \\
			&                                                       & Sn\_3 & 0.353                                             & 0.4912 & 0.4144                                            & 0.4252\\ 		\hline
	\end{tabular}}
	\label{tab1}
\end{table}

In the second time period that was selected, the parameters obtained were the following;
$$\begin{array}{lrrrrrr}
	\text{MM-gBm}&:&\mu=0.0012,&\sigma=0.0210,&p=0.6984,&q=0.4722&\\
	\text{gBm}&:&\mu=0.0004,&\sigma=0.0124&&&\\
	\text{MM-XOU}&:&\gamma=0.0016,&\sigma=0.0204,&\phi=2.6359,&p=0.6984,&q=0.4722\\
	\text{XOU}&:&\gamma=0.0014,&\sigma=0.0124,&\phi=3.0075
\end{array}
$$

Comparing the parameters obtained from data with repetitions using the XOU and those from  cleaned data (data without repetitions) using the MM-XOU, the volatility $\sigma$ from the latter model is higher. 
This is  because  the XOU is averaging apparent zero move days with higher move days and the blended average pulls down.  Due to the same reason, the $\sigma$ from MM-gBm is higher than that from gBm. 

\begin{table}[htbp]\scriptsize \centering
	\caption{MAPE  at different time periods while modelling SBU stock prices before COVID 19 in the second time period that was selected ($ 1/9/2018 - 7/25/2018 $)} 
	{\begin{tabular}{ccccccc}
			\hline 	
			\multicolumn{3}{c}{}                                                                                                                               & \multicolumn{4}{c}{\text{MAPE (\%) from different simulations}}\\ \hline
			\text{Date}        &\text{Total  days }& \text{Sn}    & \text{MM- gBm}      & \text{gBm}          &\text{ MM- XOU}      & \text{XOU}   \\
			\hline	\multirow{3}{*}{\begin{tabular}[c]{@{}c@{}}1/9/2018\\ -\\ 7/25/2018\end{tabular}} & \multirow{3}{*}{100} & Sn\_1 & 8.9709       & 8.6427      & 12.2485      & 11.9744      \\
			&                      & Sn\_2 & 8.4806       & 8.3862      & 12.5759      & 11.7782      \\
			&                      & Sn\_3 & 8.8614       & 8.9123      & 12.0063      & 12.1271      \\ \hline
			\multirow{3}{*}{\begin{tabular}[c]{@{}c@{}}1/9/2018\\ -\\ 4/18/2018\end{tabular}} & \multirow{3}{*}{50}  & Sn\_1 & 3.1406       & 6.1889      & 7.7825       & 8.1983       \\
			&                      & Sn\_2 & 3.1832       & 6.3522      & 7.874        & 8.2144       \\
			&                      & Sn\_3 & 3.0348       & 6.1124      & 7.7053       & 7.9197       \\ \hline
			\multirow{3}{*}{\begin{tabular}[c]{@{}c@{}}1/9/2018\\ -\\ 3/7/2018\end{tabular}}  & \multirow{3}{*}{30}  & Sn\_1 & 2.7566       & 5.3715      & 6.7172       & 6.4445       \\
			&                      & Sn\_2 & 2.6821       & 5.4129      & 6.4118       & 6.5325       \\
			&                      & Sn\_3 & 2.6851       & 5.3734      & 6.5541       & 6.6024     \\
			\hline
	\end{tabular}}
	\label{tab2}
\end{table}

Using the parameters from the four models in each time period,
different simulations (using different random numbers) to model the stock prices  were conducted and the forecasting accuracy of the  models was obtained using the MAPE in Equation~\eqref{eqn15}. 

\begin{figure}[h]
	\centering
	\includegraphics[width=11cm]{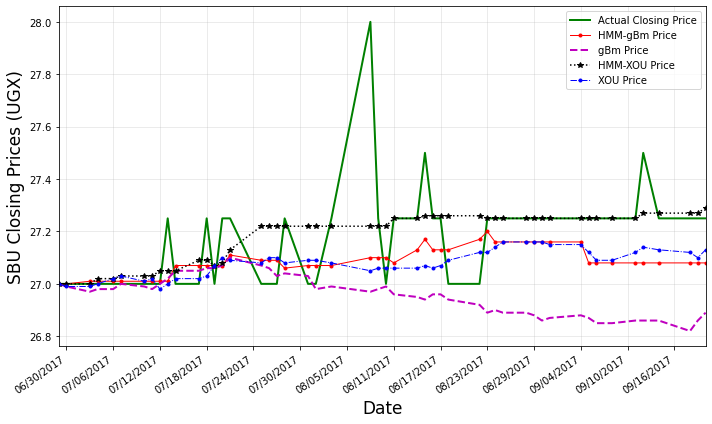} 
	\caption{Modelling stock prices of SBU before COVID 19  from 6/29/2017 to 9/20/2017 with MAPE values; 0.2413\%, 0.8487\%, 0.3202\% and 0.4872\% for MM-gBm, gBm, MM-XOU and XOU respectively.}
	\label{fig3}
\end{figure}

For the modelled 100 days in the first time period ($ 6/29/2017 - 12/2/2017 $), MM-XOU and XOU MAPE values were lower than those of MM-gBm and gBm (but close to them). For the 50 days that were modelled in the same time period ($ 6/29/2017 - 9/20/2017 $), the MM-gBm outperformed the other models, gBm being the worst. Figure~\ref{fig3} shows the forecasted values from one of the runs of each model and the actual values from $ 6/29/2017 - 9/20/2017 $. A Kolmogorov-Smirnov test on the distribution of the returns from the forecasted values of the four models  in Figure~\ref{fig3} and the actual returns in the same time period was conducted,  results were summarized in Table~\ref{tab3}.  Decisions on the distribution of the returns from each model were done based on the p-value; the probability of observing the test statistic. The distributions of the returns were similar for all the four models in this case.

\begin{table}[h]\small \centering
	\caption{Kolmogorov-Smirnov test on actual and modelled returns of DFCU  before COVID 19 (significance level$=0.01$) }
	{
		\begin{tabular}{ccccc}
			\hline
			& \text{MM-gBm }  & \text{gBm}     & \text{MM-XOU}  & \text{XOU}     \\ \hline
			p-value                                                                                                                    & 0.2611 & 0.0632  & 0.2611 & 0.1700\\ 
			Decision  & Similar   & Similar & Similar  & Similar \\ \hline
	\end{tabular}}
	
	\label{tab3}
\end{table}

When the modelled days reduced to 30, the MM-gBm still outperformed the other models, though their MAPE values were close to each other.
For the second time period modelled ($ 1/9/2018-7/25/2018 $), the MM-gBm had the lowest MAPE values for the 100, 50, 30 days that were modelled, hence outperformed the other three models. The second best model was gBm, and the MAPE values for MM-XOU and XOU were too close to each other for the time intervals chosen. Generally, the MM-gBm had  very small MAPE values when used to forecast a short time interval (30 and 50 days).

\subsection{Modelling SBU closing prices during COVID 19}
Two different time periods were modelled during the pandemic, $  1/7/2020 - 8/21/2020 $ and $  8/28/2020 - 3/15/2021 $.  Data for calibration of the models in the first time period started from $ 2/12/2019 - 1/6/2020 $ (a total of 200 days), and for the second time period, calibration data started from $ 8/12/2019 - 8/27/2020  $ (a total of 200 days) for all the four models.  All these 200 days were utilized to calibrate gBm and XOU.  In the first time period, 120 days remained after cleaning the data and were used to calibrate MM-XOU and MM-gBm, while in the second time period, 101 days remained. The  MAPE values from different simulations are shown in Table~\ref{tab4}. 
The parameters obtained were the following;
$$\begin{array}{lrrrrrr}
	\text{MM-gBm}&:&\mu=-0.0012,&\sigma=0.0202,&p=0.6625,&q=0.7797&\\
	\text{gBm}&:&\mu=-0.0007,&\sigma=0.0156&&&\\
	\text{MM-XOU}&:&\gamma=0.002,&\sigma=0.02,&\phi=4.0098,&p=0.6625,&q=0.7797\\
	\text{XOU}&:&\gamma=0.0023,&\sigma=0.0155,&\phi=3.7008
\end{array}
$$

\begin{table}[h]\scriptsize \centering
	\caption{MAPE  at different time periods while modelling SBU stock prices during COVID 19 in the first time period selected $(1/7/2020-8/21/2020)$.}
	{\begin{tabular}{ccccccc}
			\hline	\multicolumn{3}{c}{}                                                                                                                               & \multicolumn{4}{c}{\text{MAPE (\%) from different simulations}}\\\hline
			\text{Date}        &\text{Total  days} & \text{Sn}    &\text{ MM- gBm}      & \text{gBm}          & \text{MM- XOU }     & \text{XOU}   \\\hline
			\multirow{3}{*}{\begin{tabular}[c]{@{}c@{}}1/7/2020\\ - \\ 8/21/2020\end{tabular}} & \multirow{3}{*}{100}                                  & Sn\_1 & 1.6068                                                                         & 1.4782 & 8.7933                                            & 9.6931 \\
			&                                                       & Sn\_2 & 1.2661                                                                         & 1.6803 & 8.089                                             & 9.4412 \\
			&                                                       & Sn\_3 & 1.5073                                                                         & 1.5206 & 8.8426                                            & 9.4373 \\\hline
			\multirow{3}{*}{\begin{tabular}[c]{@{}c@{}}1/7/2020\\ -\\ 3/30/2020\end{tabular}}  & \multirow{3}{*}{50}                                   & Sn\_1 & 0.9703                                                                         & 1.6972 & 6.0762                                            & 6.4011 \\
			&                                                       & Sn\_2 & 0.6944                                                                         & 1.7961 & 5.5214                                            & 6.0274 \\
			&                                                       & Sn\_3 & 0.8392                                                                         & 1.6972 & 5.6442                                            & 6.0478 \\\hline
			\multirow{3}{*}{\begin{tabular}[c]{@{}c@{}}1/7/2020\\ -\\ 2/18/2020\end{tabular}}  & \multirow{3}{*}{30}                                   & Sn\_1 & 0.8218                                                                         & 1.7467 & 4.0004                                            & 4.1578 \\
			&                                                       & Sn\_2 & 0.8565                                                                         & 1.666  & 3.8404                                            & 4.4224 \\
			&                                                       & Sn\_3 & 0.864                                                                          & 1.8422 & 4.1519                                            & 4.3761
			\\ 		\hline
	\end{tabular}}
	\label{tab4}
\end{table}

The parameters obtained after calibration of the four models in the second time period selected were the following;
$$\begin{array}{lrrrrrr}
	\text{MM-gBm}&:&\mu=-0.0019,&\sigma=0.0198,&p=0.8061,&q=0.81&\\
	\text{gBm}&:&\mu=-0.001,&\sigma=0.014&&&\\
	\text{MM-XOU}&:&\gamma=0.0007,&\sigma=0.0196,&\phi=6.3790,&p=0.8061,&q=0.81\\
	\text{XOU}&:&\gamma=-0.5493,&\sigma=1.1478,&\phi=3.7485
\end{array}
$$

In Table~\ref{tab5} the MAPE values that were obtained for the different time periods selected in the second time period during COVID 19 are shown.

\begin{table}[htbp]\scriptsize \centering
	\caption{MAPE  at different time periods while modelling SBU stock prices during COVID 19 in the second time period selected ($8/28/2020-3/15/2021$).\label{tab5}}
	
	\begin{tabular}{ccccccc}
		\hline 	\multicolumn{3}{c}{}                                                                                                                               & \multicolumn{4}{c}{\text{MAPE (\%) from different simulations}}\\\hline
		\text{Date}        &\text{Total  days }& \text{Sn}    & \text{MM- gBm }     & \text{gBm}          & \text{MM- XOU  }    & \text{XOU}   \\\hline
		\multirow{3}{*}{\begin{tabular}[c]{@{}c@{}}8/28/2020\\ -\\ 3/15/2021\end{tabular}}  & \multirow{3}{*}{100}                                 & Sn\_1 & 3.2959                                                                         & 3.6994 & 6.5813                                            & $\infty$ \\
		&                                                      & Sn\_2 & 3.2761                                                                         & 3.582  & 6.0679                                            & $\infty$ \\
		&                                                      & Sn\_3 & 3.9818                                                                         & 3.8271 & 5.9668                                            & $\infty$ \\\hline
		\multirow{3}{*}{\begin{tabular}[c]{@{}c@{}}8/28/2020\\ -\\ 11/19/2020\end{tabular}} & \multirow{3}{*}{50}                                  & Sn\_1 & 0.8196                                                                         & 1.399  & 4.5341                                            & $\infty$ \\
		&                                                      & Sn\_2 & 0.7231                                                                         & 1.3648 & 4.1866                                            & $\infty$ \\
		&                                                      & Sn\_3 & 0.6926                                                                         & 1.3699 & 4.0253                                            & $\infty$ \\\hline
		\multirow{3}{*}{\begin{tabular}[c]{@{}c@{}}8/28/2020\\ -\\ 10/14/2020\end{tabular}} & \multirow{3}{*}{30}                                  & Sn\_1 & 0.5708                                                                         & 1.2333 & 1.1476                                            & $\infty$ \\
		&                                                      & Sn\_2 & 0.5505                                                                         & 1.5022 & 1.339                                             & $\infty$ \\
		&                                                      & Sn\_3 & 0.571                                                                          & 1.3287 & 1.6371                                            & $\infty$
		\\ 		\hline
	\end{tabular}
	
\end{table}

In the first time period, the $\sigma$ parameter from MM-gBm is greater than that from gBm and the one from MM-XOU is greater than that from XOU. However, for the second time period the $\sigma$ for gBm was smaller than the one for MM-gBm while that of XOU was higher than that of MM-XOU.
In the first time period, the MM-gBm had the smallest MAPE values of all the four models followed by gBm for the 100 forecasted days ($ 1/7/2020 - 8/21/2020 $).  MM-XOU performed better than XOU for all the simulations. As the number of days modelled reduced to 50, the MM-gBm still had the lowest MAPE values compared to those of the other three models, which were lower than those obtained when the days modelled were 100.

\begin{figure}[htbp]
	\centering
	
	\includegraphics[width=11cm]{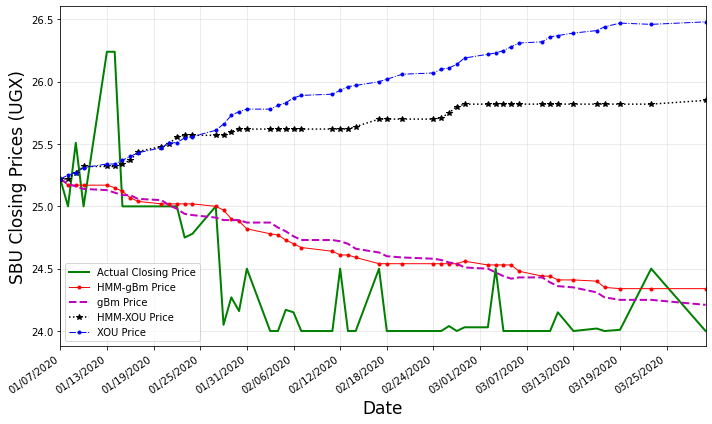} 
	
	\caption{Modelling stock prices of SBU during COVID 19  from 1/7/2020 to 3/30/2020 with MAPE values; 0.8407\%, 1.6972\%, 5.3168\% and 6.4011\% for MM-gBm, gBm, MM-XOU and XOU respectively.}
	\label{fig4}
\end{figure}

The MAPE values for the other models while modelling 50 days were also smaller than those obtained while modelling 100 days in this time period. However, the MAPE values when modelling 30 days ($ 1/7/2020 - 2/18/2020 $) using MM-gBm and gBm were close to those obtained while modelling 50 days, and there was a significant difference between the MAPE values of the XOU and MM-gBm when the modelled days reduced to 30. Figure~\ref{fig4} shows forecasted values of SBU from the four models from $ 1/7/2020-3/30/2020 $. A Kolmogorov-Smirnov test on the distribution of the returns from the forecasted values of the four models  in Figure~\ref{fig4} and the actual returns in the same time period was conducted, results were summarized in Table~\ref{tab6}.  The distributions of the returns were different for gBm and XOU, and  similar for MM-gBm 
and MM-XOU.

\begin{table}[h]\small \centering
	\caption{Kolmogorov-Smirnov test on actual and modelled returns of SBU  during COVID 19 (significance level$=0.01$) }
	{
		\begin{tabular}{ccccc}
			\hline
			& \text{MM-gBm}   & \text{gBm}     & \text{MM-XOU}  & \text{XOU}     \\ \hline
			p-value                                                                                                                                                                             & 0.0198 & $3.61e^{-6}$  & 0.0133 & $2.69e^{-8}$ \\ 
			Decision  & Similar   & Different & Similar  & Different  \\ \hline
	\end{tabular}}
	\label{tab6}
\end{table}

The mean-reverting speed in the second time period modelled was negative for the XOU and  unable to model the prices of SBU in this time period.
It was diverging away from the mean hence,  forecasting extremely large values (infinity). For the 100 days modelled from 8/28/2020 to 3/15/2021, the MAPE values of MM-gBm, gBm and MM-XOU were close to each other. When the days reduced 50 and 30, the MM-gBm had the smallest MAPE values compared to the other two models, with those from 30 days being smaller than those from 50 days. The MM-gBm and gBm models performed better than the MM-XOU for all the modelled time periods.

\subsection{Modelling SBU prices after COVID 19}
The two different time periods were modelled aftre the pandemic were $ 7/17/2023-2/1/2024 $ and $ 1/4/2024 - 7/22/2024 $.  Data for calibration of the models in the first time period started from $ 7/8/2022-7/11/2023,$ and for the second time period, calibration data started from $ 11/17/2022-1/2/2024 $ for all the four models. All these days were used to get the parameters of XOU and gBm.  All these 200 days were utilized to calibrate gBm and XOU.  In the first time period, 73 days remained after cleaning the data and were used to calibrate MM-XOU and MM-gBm, while in the second time period, 69 days remained.  The MAPE values from three different simulations are shown in Table~\ref{tab7} for each model in each time period. 
The parameters obtained were the following;
$$\begin{array}{lrrrrrr}
	\text{MM-gBm}&:&\mu=0.0027,&\sigma=0.0426,&p=0.8254,&q=0.6806&\\
	\text{gBm}&:&\mu=0.0010,&\sigma=0.0255&&&\\
	\text{MM-XOU}&:&\gamma=0.0086,&\sigma=0.0420,&\phi=2.8428,&p=0.8254,&q=0.6806\\
	\text{XOU}&:&\gamma=0.0073,&\sigma=0.0254,&\phi=2.9878
\end{array}
$$

\begin{table}[h]\scriptsize \centering
	\caption{MAPE  at different time periods while modelling SBU stock prices after COVID 19 in the first time period selected $7/17/2023- 2/1/2024$.}
	{\begin{tabular}{ccccccc}
			\hline 	\multicolumn{3}{c}{}                                                                                                                               & \multicolumn{4}{c}{\text{MAPE (\%) from different simulations}}\\\hline
			\text{Date}        &\text{Total  days} & \text{Sn}    & \text{MM- gBm }     & \text{gBm}          & \text{MM- XOU}      & \text{XOU}   \\\hline
			\multirow{3}{*}{\begin{tabular}[c]{@{}c@{}}7/17/2023\\ - \\ 2/1/2024\end{tabular}} & \multirow{3}{*}{100}                                  & Sn\_1 & 9.9991                                                                         & 9.1822  & 17.6859                                           & 19.6529 \\
			&                                                       & Sn\_2 & 9.923                                                                          & 9.9292  & 17.0106                                           & 19.8711 \\
			&                                                       & Sn\_3 & 9.3594                                                                         & 10.1696 & 17.0381                                           & 19.6564 \\\hline
			\multirow{3}{*}{\begin{tabular}[c]{@{}c@{}}7/17/2023\\ -\\ 10/22/2023\end{tabular}}  & \multirow{3}{*}{50}                                   & Sn\_1 & 3.3004                                                                         & 6.3584  & 10.1344                                           & 11.3321 \\
			&                                                       & Sn\_2 & 3.3179                                                                         & 6.2759  & 10.1346                                           & 11.6493 \\
			&                                                       & Sn\_3 & 3.0445                                                                         & 6.5199  & 10.5218                                           & 11.3823 \\\hline
			\multirow{3}{*}{\begin{tabular}[c]{@{}c@{}}7/17/2023\\ -\\ 9/10/2023\end{tabular}}   & \multirow{3}{*}{30}                                   & Sn\_1 & 0.998                                                                          & 1.9113  & 3.8833                                            & 4.5329  \\
			&                                                       & Sn\_2 & 0.8969                                                                         & 1.7456  & 3.4029                                            & 4.5828  \\
			&                                                       & Sn\_3 & 0.8185                                                                         & 2.0913  & 2.565                                             & 4.6964 
			\\ 		\hline
	\end{tabular}}
	\label{tab7}
\end{table}

The parameters obtained in the second time period selected after COVID 19 were the following;
$$\begin{array}{lrrrrrr}
	\text{MM-gBm}&:&\mu=0.0062,&\sigma=0.0392,&p=0.8168,&q=0.6567&\\
	\text{gBm}&:&\mu=0.0021,&\sigma=0.0230&&&\\
	\text{MM-XOU}&:&\gamma=0.0035,&\sigma=0.0387,&\phi=1.7247,&p=0.8168,&q=0.6567\\
	\text{XOU}&:&\gamma=0.0041,&\sigma=0.0229,&\phi=2.7758
\end{array}
$$

Table~\ref{tab8} shows the MAPE values from different time periods selected in the second time period after COVID 19.

\begin{table}[h]\scriptsize \centering
	\caption{MAPE  at different time periods while modelling SBU stock prices after COVID 19 for the second time period selected ($1/4/2024-7/22/2024$)} 
	{
		\begin{tabular}{ccccccc}
			\hline 	\multicolumn{3}{c}{}                                                                                                                               & \multicolumn{4}{c}{\text{MAPE (\%) from different simulations}}\\\hline
			\text{Date}        &\text{Total  days} & \text{Sn}    & \text{MM- gBm }     & \text{gBm}          & \text{MM- XOU }     & \text{XOU}   \\\hline	\multirow{3}{*}{\begin{tabular}[c]{@{}c@{}}1/4/2024\\ -\\ 7/22/2024\end{tabular}} & \multirow{3}{*}{100}                                 & Sn\_1 & 2.683                                             & 2.7376 & 14.0812                                           & 17.8054 \\
			&                                                      & Sn\_2 & 2.0438                                            & 3.1939 & 14.5761                                           & 17.2657 \\
			&                                                      & Sn\_3 & 2.4448                                            & 3.2292 & 14.207                                            & 17.6772 \\\hline
			\multirow{3}{*}{\begin{tabular}[c]{@{}c@{}}1/4/2024\\ -\\ 4/17/2024\end{tabular}} & \multirow{3}{*}{50}                                  & Sn\_1 & 1.3185                                            & 3.3877 & 5.554                                             & 7.8168  \\
			&                                                      & Sn\_2 & 1.3767                                            & 3.1256 & 6.9464                                            & 7.7757  \\
			&                                                      & Sn\_3 & 1.0229                                            & 3.0608 & 5.477                                             & 8.2751  \\\hline
			\multirow{3}{*}{\begin{tabular}[c]{@{}c@{}}1/4/2024\\ -\\ 3/6/2024\end{tabular}}  & \multirow{3}{*}{30}                                  & Sn\_1 & 0.8077                                            & 3.0603 & 2.2261                                            & 3.8501  \\
			&                                                      & Sn\_2 & 0.8184                                            & 3.2635 & 2.4325                                            & 3.9943  \\
			&                                                      & Sn\_3 & 0.8595                                            & 2.9187 & 2.4598                                            & 3.7694 
			\\ 		\hline
	\end{tabular}}\label{tab8}
\end{table}

\begin{figure}[htbp]
	\centering
	\includegraphics[width=11cm]{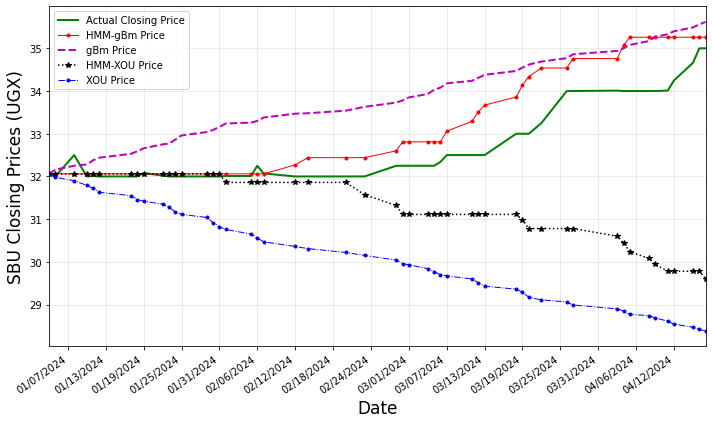} 
	\caption{Modelling stock prices of SBU after COVID 19  from 1/4/2024 to 4/17/2024  with MAPE values; 0.7426\%, 3.3877\%, 4.3403\% and 7.8168\% for MM-gBm, gBm, MM-XOU and XOU respectively.}
	\label{fig5}
\end{figure}

The volatility parameter from the non-MM models was always smaller than that from the MM-gBm and MM-XOU. For all the time periods modelled, the MM-gBm had the smallest MAPE values, with those from a short time modelled period being the smallest. All MAPE values for the 30 modelled days while using MM-gBm were below 1\%, indicating that this model performs so well while modelling a short time period. The second performing model was gBm for all the time periods, with smaller MAPE values compared to those of MM-XOU and XOU. 
However, the MAPE values of MM-XOU and XOU were close to each other for all the 100 and 50 days modelled and slightly far from each other for 30 modelled days in each time period. Figure~\ref{fig5} shows the forecasted values of SBU prices after the pandemic from $ 1/4/2024 - 4/17/2024 $ for one of the simulations in each case with 0.7426\%, 3.3877\%, 4.3403\% and 7.8168\% for MM-gBm, gBm, MM-XOU and XOU respectively.

A Kolmogorov-Smirnov test on the distribution of the returns from the forecasted values of the four models  in Figure~\ref{fig5} and the actual returns in the same time period was conducted, results were summarized in Table~\ref{tab9}.  The distributions of the returns were different while using gBmand XOU, and similar while using MM-gBm and MM-XOU.
Generally while modelling the closing daily stock prices,  MM-gBm captures the market scenario of the SBU price movements very well compared to gBm, MM-XOU and XOU especially when a short time period of 30 days is modelled.

\begin{table}[h]\small \centering
	\caption{Kolmogorov-Smirnov test on actual and modelled returns of SBU  after COVID 19 (significance level$=0.01$)}
	{
		\begin{tabular}{ccccc}
			\hline
			& \text{MM-gBm}   & \text{gBm}     & \text{MM-XOU}  & \text{XOU}     \\ \hline
			p-value                                                                                         & 0.53552 & $4.22e^{-13}$  & 0.0103 & $5.33e^{-21}$  \\ 
			Decision & Similar   & Different & Similar  & Different  \\ \hline
	\end{tabular}}
	\label{tab9}
\end{table}

\subsection{Modelling BOBU closing  prices after COVID 19}

We modelled the closing daily prices of BOBU after COVID 19 for two time periods and the results obtained were similar to those obtained while modelling SBU prices before and after COVID 19. The two time periods that were modelled were; $ 10/2/2023-6/19/2024 $ and $ 5/29/2024-12/5/2024 $. Data from $ 8/31/2021-9/29/2023 $ and $ 8/1/2022-5/27/2024 $ (200 days)  was utilized for calibration of the  models in the first time period and the second time period respectively.   All these 200 days were utilized to calibrate gBm and XOU. 
In the first time period, 60 days remained after cleaning the data and were used to calibrate MM-XOU and MM-gBm, while in the second time period, 43 days remained. The different MAPE values from different simulations while using each of the models are summarized in Table~\ref{tab10}. 
The parameters obtained were the following;
$$\begin{array}{lrrrrrr}
	\text{MM-gBm}&:&\mu=-0.0045,&\sigma=0.3143,&p=0.8286,&q=0.6034&\\
	\text{gBm}&:&\mu=-0.0013,&\sigma=0.1701,&&&\\
	\text{MM-XOU}&:&\gamma=0.2016,&\sigma=0.3248,&\phi=3.0725,&p=0.8286,&q=0.6034\\
	\text{XOU}&:&\gamma=0.0847,&\sigma=0.1729,&\phi=2.9811
\end{array}
$$

\begin{table}[h]\scriptsize \centering
	\caption{MAPE  at different time periods while modelling BOBU stock prices after COVID 19 in the first time period ($10/2/2023-6/19/2024$)}
	{
		\begin{tabular}{ccccccc}
			\hline 	\multicolumn{3}{c}{}                                                                                                                               & \multicolumn{4}{c}{\text{MAPE (\%) from different simulations}}\\\hline
			\text{Date}        &\text{Total  days} & \text{Sn}    & \text{MM- gBm}      & \text{gBm}          & \text{MM- XOU}      & \text{XOU}   \\\hline
			\multirow{3}{*}{\begin{tabular}[c]{@{}c@{}}10/2/2023\\ - \\ 6/19/2024\end{tabular}} & \multirow{3}{*}{100}                                  & Sn\_1 & 8.168                                                                          & 7.5276 & 14.8915                                           & 14.1142 \\
			&                                                       & Sn\_2 & 8.5414                                                                         & 9.8905 & 14.5793                                           & 14.3033 \\
			&                                                       & Sn\_3 & 7.3712                                                                         & 7.843  & 15.0762                                           & 15.4434 \\\hline
			\multirow{3}{*}{\begin{tabular}[c]{@{}c@{}}10/2/2023\\ -\\ 2/9/2024\end{tabular}}   & \multirow{3}{*}{50}                                   & Sn\_1 & 1.5232                                                                         & 1.6039 & 16.3712                                           & 16.4052 \\
			&                                                       & Sn\_2 & 1.8886                                                                         & 2.4545 & 16.9256                                           & 17.6227 \\
			&                                                       & Sn\_3 & 1.7245                                                                         & 1.8517 & 15.0771                                           & 16.4017 \\\hline
			\multirow{3}{*}{\begin{tabular}[c]{@{}c@{}}10/2/2023\\ -\\ 12/12/2023\end{tabular}} & \multirow{3}{*}{30}                                   & Sn\_1 & 0.8643                                                                         & 1.8451 & 12.0242                                           & 13.5617 \\
			&                                                       & Sn\_2 & 0.7361                                                                         & 1.6706 & 11.9708                                           & 14.8707 \\
			&                                                       & Sn\_3 & 0.6238                                                                         & 1.742  & 13.4896                                           & 14.5617
			\\ 		\hline
	\end{tabular}}
	\label{tab10}
\end{table}
The parameters obtained in the second time period of BOBU atfer COVID 19 were the following;
$$\begin{array}{lrrrrrr}
	\text{MM-gBm}&:&\mu=0.0070,&\sigma=0.3723,&p=0.8846,&q=0.5714&\\
	\text{gBm}&:&\mu=0.00147,&\sigma=0.1694&&&\\
	\text{MM-XOU}&:&\gamma=0.2084,&\sigma=0.3826,&\phi=3.1142,&p=0.8846,&q=0.5714\\
	\text{XOU}&:&\gamma=0.0876,&\sigma=0.1723,&\phi=2.9279
\end{array}
$$

\begin{figure}[h]
	\centering
	\includegraphics[width=11cm]{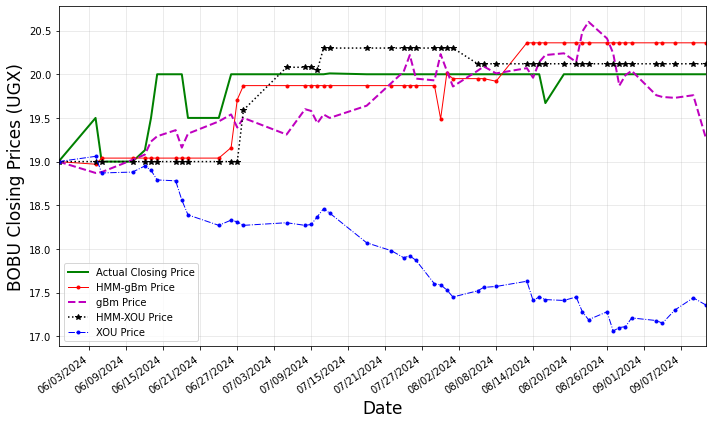} 
	
	\caption{Modelling stock prices of BOBU after COVID 19  from 5/29/2024 to 9/11/2024  with MAPE values; 0.7726\%, 1.4334\%, 1.4016\% and 9.8524\% for MM-gBm, gBm, MM-XOU and XOU respectively.}
	\label{fig6}
\end{figure}

The volatility $\sigma$ for non-MM models was always smaller  than that of the MM-gBm and MM-XOU. 	
The MAPE values obtained in the second time period are shown in Table~\ref{tab11}. 
For all the time periods modelled, the MM-gBm outperformed all the other three models, with the smallest MAPE values. The MAPE values for the gBm were close to those of the MM-gBm and it is the second best performing model after MM-gBm for all the modelled time periods. All MAPE values in the first time period for MM-XOU and XOU were greater than 10\% with those for 100 days being the highest. Figure~\ref{fig6} shows forecasted values of the four models from $ 5/29/2024 - 9/11/2024 $.

\begin{table}[h]\scriptsize \centering
	\caption{MAPE  at different time periods while modelling BOBU stock prices after COVID 19 for the second time period selected ($5/29/2024-12/5/2024$).}
	{\begin{tabular}{ccccccc}
			\hline 	\multicolumn{3}{c}{}                                                                                                                               & \multicolumn{4}{c}{\text{MAPE (\%) from different simulations}}\\\hline
			\text{Date}        &\text{Total  days} & \text{Sn}    & \text{MM- gBm}      & \text{gBm}          & \text{MM- XOU }     & \text{XOU}   \\\hline
			\multirow{3}{*}{\begin{tabular}[c]{@{}c@{}}5/29/2024\\ -\\ 12/5/2024\end{tabular}} & \multirow{3}{*}{100}                                 & Sn\_1 & 2.9236                                                                         & 3.3892 & 7.4397                                            & 15.8958 \\
			&                                                      & Sn\_2 & 4.2533                                                                         & 4.219  & 6.158                                             & 15.499  \\
			&                                                      & Sn\_3 & 3.1015                                                                         & 4.2417 & 7.5835                                            & 15.6516 \\\hline
			\multirow{3}{*}{\begin{tabular}[c]{@{}c@{}}5/29/2024\\ -\\ 9/11/2024\end{tabular}} & \multirow{3}{*}{50}                                  & Sn\_1 & 1.0268                                                                         & 1.4334 & 2.6261                                            & 9.8524  \\
			&                                                      & Sn\_2 & 0.9887                                                                         & 1.5002 & 1.8659                                            & 10.7443 \\
			&                                                      & Sn\_3 & 1.7681                                                                         & 2.402  & 1.9662                                            & 10.1511 \\\hline
			\multirow{3}{*}{\begin{tabular}[c]{@{}c@{}}5/29/2024\\ -\\ 8/5/2024\end{tabular}}  & \multirow{3}{*}{30}                                  & Sn\_1 & 0.9767                                                                         & 0.9932 & 1.1626                                            & 10.0355 \\
			&                                                      & Sn\_2 & 0.8521                                                                         & 1.4083 & 1.4955                                            & 8.4843  \\
			&                                                      & Sn\_3 & 0.9711                                                                         & 1.1465 & 1.3022                                            & 9.5225 
			\\ 		\hline
	\end{tabular}}
	\label{tab11}
\end{table}

A Kolmogorov-Smirnov test on the distribution of the returns from the forecasted values of the four models  in Figure~\ref{fig6} and the actual returns in the same time period was conducted, results were shown in Table~\ref{tab12}. The distributions were similar when using MM-gBm and MM-XOU, and they were different for the non-MM models; gBm and XOU.

\begin{table}[h]\small \centering
	\caption{Kolmogorov-Smirnov test on actual and modelled returns of BOBU  after COVID 19 (significance level$=0.01$) }
	{
		\begin{tabular}{ccccc}
			\hline
			& \text{MM-gBm }  & \text{gBm}     & \text{MM-XOU}  & \text{XOU}     \\ \hline
			p-value                                                                                                 & 0.9999& $7.99e^{-5}$  & 0.9973 & $3.61e^{-6}$ \\ 
			Decision  & Similar   & Different & Similar  & Different  \\ \hline
	\end{tabular}}
	\label{tab12}
\end{table}

\subsection{Modelling ALSIUG closing daily prices after COVID 19}

ALSIUG is the main stock market index in Uganda which tracks the performance of companies listed at USE. Due to this fact, we also applied the four models on the ALSIUG data which is not constant much of the time, but rather moves up and down all the time. 
Very few repetitions were found in the data for ALSIUG hence, the parameters of the MM-XOU and XOU models are equal or close to each other.  The two time periods that were modelled were; $ 7/15/2024-12/2/2024 $ and $ 11/2/2023-3/28/2024 $. Data from $ 9/20/2023-7/12/2024 $ and $ 1/12/2023-11/1/2023 $ (200 days)  was utilized for calibration of the  models in the first time period and the second time period respectively.   All these 200 days were utilized to calibrate gBm and XOU. 
In the first time period, 199 days remained after cleaning the data and were used to calibrate MM-XOU and MM-gBm, while in the second time period, 198 days remained. The  different MAPE values in the first time period from different simulations while using each of the models are summarized in Table~\ref{tab13}. 
The parameters obtained were the following;

$$\begin{array}{lrrrrrr}
	\text{MM-gBm}&:&\mu=0.0005,&\sigma=0.0171,&p=0.00,&q=0.9949&\\
	\text{gBm}&:&\mu=0.0005,&\sigma=0.0170&&&\\
	\text{MM-XOU}&:&\gamma=0.00075,&\sigma=0.016699,&\phi=6.3208,&p=0.00,&q=0.9949\\
	\text{XOU}&:&\gamma=0.00076,&\sigma=0.16947,&\phi=6.3234
\end{array}
$$

\begin{table}[h]\scriptsize \centering
	\caption{MAPE  at different time periods while modelling ALSIUG stock prices after COVID 19 for the first time period selected ($7/15/2024-12/2/2024$).}
	{\begin{tabular}{ccccccc}
			\hline 	\multicolumn{3}{c}{}                                                                                                                               & \multicolumn{4}{c}{\text{MAPE (\%) from different simulations}}\\\hline
			\text{Date}        &\text{Total  days} & \text{Sn}    & \text{MM- gBm}      & \text{gBm}          & \text{MM- XOU}      & \text{XOU}   \\\hline
			\multirow{3}{*}{\begin{tabular}[c]{@{}c@{}}7/15/2024\\ - \\ 12/2/2024\end{tabular}} & \multirow{3}{*}{100} & Sn\_1 & 3.7187        & 3.8423      & 6.9716       & 6.8776      \\
			&                      & Sn\_2 & 3.5496        & 3.6602      & 6.3187       & 6.8611      \\
			&                      & Sn\_3 & 3.5783        & 3.5489      & 6.5541       & 6.6365      \\ \hline
			\multirow{3}{*}{\begin{tabular}[c]{@{}c@{}}7/15/2024\\ -\\ 9/20/2024\end{tabular}}  & \multirow{3}{*}{50}  & Sn\_1 & 1.7136        & 1.7107      & 1.6387       & 1.6517      \\
			&                      & Sn\_2 & 1.6844        & 1.4741      & 1.5531       & 1.5373      \\
			&                      & Sn\_3 & 1.7618        & 1.7354      & 1.6285       & 1.6106      \\ \hline
			\multirow{3}{*}{\begin{tabular}[c]{@{}c@{}}7/15/2024\\ -\\ 8/23/2024\end{tabular}}  & \multirow{3}{*}{30}  & Sn\_1 & 1.8439        & 1.9833      & 0.9067       & 1.0107      \\
			&                      & Sn\_2 & 1.9061        & 1.9876      & 0.8314       & 1.0684      \\
			&                      & Sn\_3 & 1.94          & 1.9833      & 1.0862       & 0.8863 	\\ 		\hline
	\end{tabular}}
	\label{tab13}
\end{table}

The parameters obtained were the following used to model ALSIUG prices in the second time period selected;

$$\begin{array}{lrrrrrr}
	\text{MM-gBm}&:&\mu=-0.00143,&\sigma=0.0171,&p=0.0,&q=0.9898&\\
	\text{gBm}&:&\mu=-0.0014,&\sigma=0.0170,&&&\\
	\text{MM-XOU}&:&\gamma=0.000334,&\sigma=0.016975,&\phi=11.5956,&p=0.0,&q=0.9898\\
	\text{XOU}&:&\gamma=0.00034,&\sigma=0.01689,&\phi=11.4491
\end{array}
$$

\begin{table}[h]\scriptsize \centering
	\caption{MAPE  at different time periods while modelling BOBU stock prices after COVID 19 for the second time period selected ($11/2/2023-3/28/2024$).}
	{
		\begin{tabular}{ccccccc}
			
			\hline 	\multicolumn{3}{c}{}                                                                                                                               & \multicolumn{4}{c}{\text{MAPE (\%) from different simulations}}\\\hline
			\text{Date}        &\text{Total  days} & \text{Sn}    & \text{MM- gBm }     & \text{gBm}          & \text{MM- XOU }     & \text{XOU}   \\\hline
			\multirow{3}{*}{\begin{tabular}[c]{@{}c@{}}11/2/2023\\ -\\ 3/28/2024\end{tabular}}  & \multirow{3}{*}{100}                                 & Sn\_1 & 5.5881                                            & 5.1703 & 9.9405                                            & 9.1726 \\
			&                                                      & Sn\_2 & 5.1516                                            & 5.33   & 9.4115                                            & 9.9684 \\
			&                                                      & Sn\_3 & 5.2844                                            & 5.5085 & 9.4864                                            & 9.767  \\\hline
			\multirow{3}{*}{\begin{tabular}[c]{@{}c@{}}11/2/2023\\ -\\ 1/15/2024\end{tabular}}  & \multirow{3}{*}{50}                                  & Sn\_1 & 1.2104                                            & 1.2805 & 6.7421                                            & 6.7227 \\
			&                                                      & Sn\_2 & 1.158                                             & 1.0473 & 6.8729                                            & 6.4583 \\
			&                                                      & Sn\_3 & 1.1068                                            & 1.0662 & 6.5799                                            & 6.3783 \\\hline
			\multirow{3}{*}{\begin{tabular}[c]{@{}c@{}}11/2/2023\\ -\\ 12/13/2023\end{tabular}} & \multirow{3}{*}{30}                                  & Sn\_1 & 0.9741                                            & 0.9875 & 4.5354                                            & 4.0847 \\
			&                                                      & Sn\_2 & 1.0653                                            & 1.0122 & 4.1961                                            & 4.224  \\
			&                                                      & Sn\_3 & 0.9392                                            & 1.0173 & 4.0111                                            & 4.2647
			\\ 		\hline
	\end{tabular}}
	\label{tab15}
\end{table}

\begin{figure}[h]
	\centering
	\includegraphics[width=11cm]{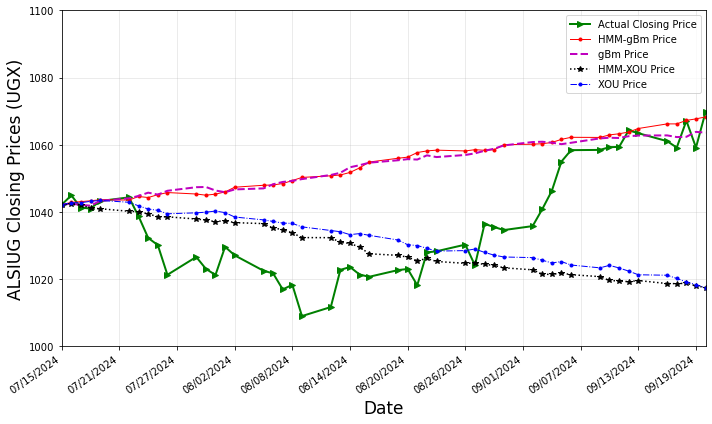} 
	\caption{Modelling stock prices of ALSIUG after COVID 19  from 7/15/2024 to 9/20/2024  with MAPE values; 1.7266\%,1.7107\%,1.6387\% and 1.6517\% for MM-gBm, gBm, MM-XOU and XOU respectively.}
	\label{fig7}
\end{figure}

The XOU and MM-XOU models have MAPE values and parameters that are close to each other. Since the data has 1 or 2  repetitions unlike the stocks with so many repeated values in just one year. The gBm and MM-gBm also have close or equal MAPE values and parameters as well.  For the 100 days modelled  in the first time period ($ 7/15/2024-12/2/2024 $), the MM-gBm and gBm performed better than the other two models. The 50 days days modelled in the first time period, all the four models performed equally well, while for the 30 days modelled, the MM-XOU and XOU had MAPE values slightly lower than those of the other models.
For the second time period, Mape values of MM-gBm and gBm were smaller than those of MM-XOU and XOU for the 100, 50 and 30 days that were modelled. The MAPE values reduced for all the models as the number of days modelled reduced. Figure~\ref{fig7} shows forecasted closing prices and actual closing prices of ALSIUG from $ 7/15/2024-9/20/2024 $ for the four models. 

A Kolmogorov-Smirnov test on the distribution of the returns from the forecasted values of the four models  in Figure~\ref{fig7} and the actual returns in the same time period was conducted, results were summarized in Table~\ref{tab15}. The distributions of the returns were different while using each of the four models. The results obtained while modelling ALSIUG prices show that the MM models utilized in this study work with illiquid stocks with constant prices much of the time.

\begin{table}[h]\small \centering
	\caption{Kolmogorov-Smirnov test on returns of ALSIUG after COVID 19 (significance level $=0.01$)}
	{\begin{tabular}{ccccc}
			\hline
			& \text{MM-gBm }  & \text{gBm}     & \text{MM-XOU}  & \text{XOU}     \\ \hline
			p-value                                                                                                             & $1.15e^{-6}$ & $1.15e^{-6}$  & $9.98e^{-8}$ & $9.98e^{-8}$ \\ 
			Decision& Different   & Different & Different  & Different \\ \hline
	\end{tabular}}\label{tab15}
\end{table}

\subsection{ MM-gBm conclusion about correlation of the stocks 
}
According to \cite{shinzawa2006effect} an extremely large window can contain a region that is more than required that more information is broadly distributed and also an extremely small window leads to high noise levels in the data.
An appropriate selection of the window size is useful in correlation analysis.  \cite{gu2024sliding} emphasizes that  a window
size that is not too small and not too large eliminates the noise in correlation computations.
\cite{keilholz2013dynamic} shows that shorter window sizes exhibit greater variances over time. \cite{gogtay2017principles} explains that small sample sizes while calculating correlation coefficients may show false positive relationships.

\subsubsection{Rolling window correlation of stocks}
Stocks at USE have a number of non trading days that are different in a given year.
So, to make sure that we had a price on each day, we filled the days for no trade with the previous day's price, and if there was no trading on the previous day, the price on the last day of trading was used as the prices for all the days with no trade. Forexample; 
if there was no trading on Friday and Thursday, and there was trading on Wednesday. The prices for Thursday and Friday was taken as that of Wednesday. This daily data that was filled for all the trading days in a year was utilized to generate weekly data in order to have a price on every week day chosen.
The weekly data was generated from the daily data by extracting data on the same day, Friday. This weekly data was used in calculations of rolling correlations of log returns.

\begin{figure}[htbp]
	\centering
	
	\includegraphics[width=10cm]{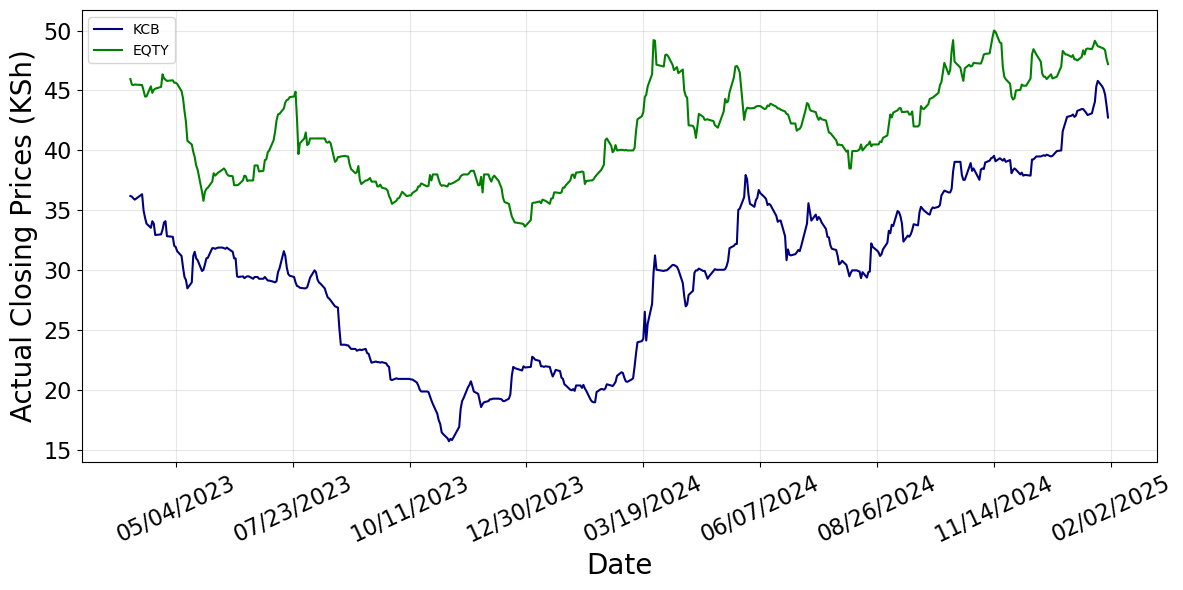} 
	\caption{Actual closing prices of KCB and EQTY.}
	\label{fig8}
\end{figure}

\begin{figure}[h]
	\centering
	\subfloat[EQTY and KCB.]{%
		\resizebox*{7cm}{!}{\includegraphics{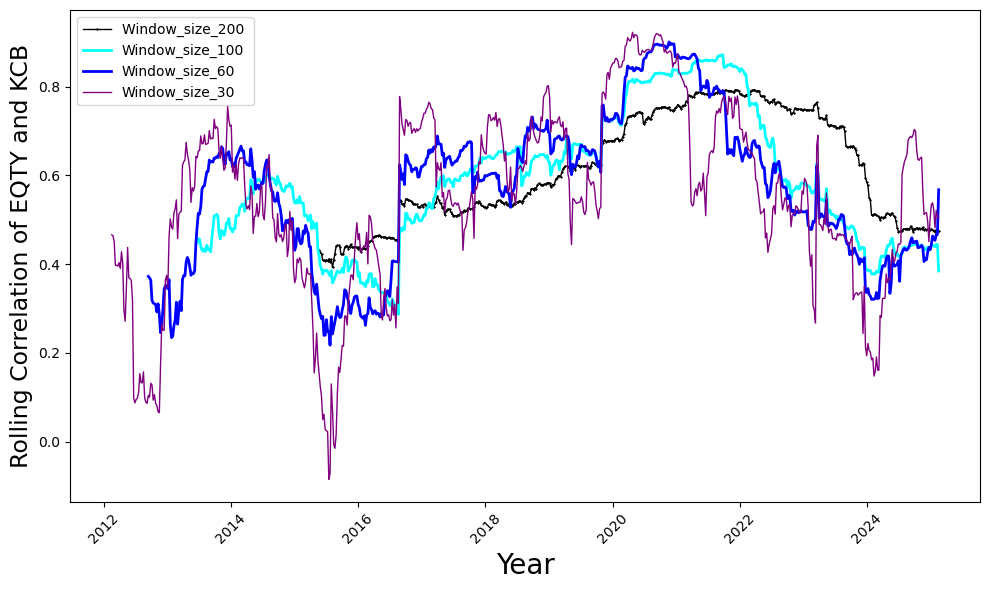}}}\hspace{5pt}
	\hspace{1cm}
	\subfloat[ALSIUG and EQTY.]{%
		\resizebox*{7cm}{!}{\includegraphics{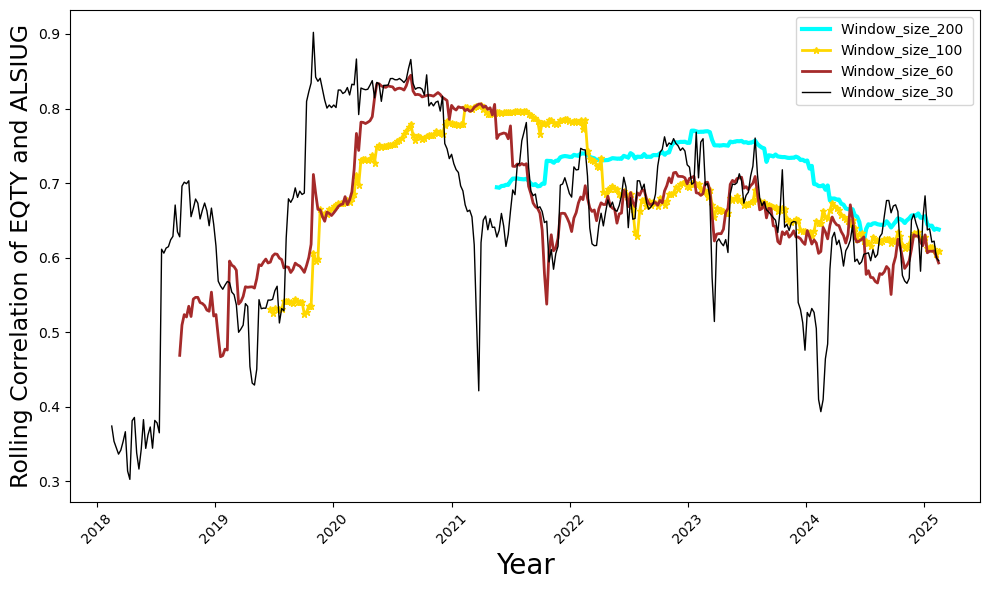}}}\hspace{5pt}
	
	\subfloat[EQTY and KCB.]{%
		\resizebox*{7cm}{!}{\includegraphics{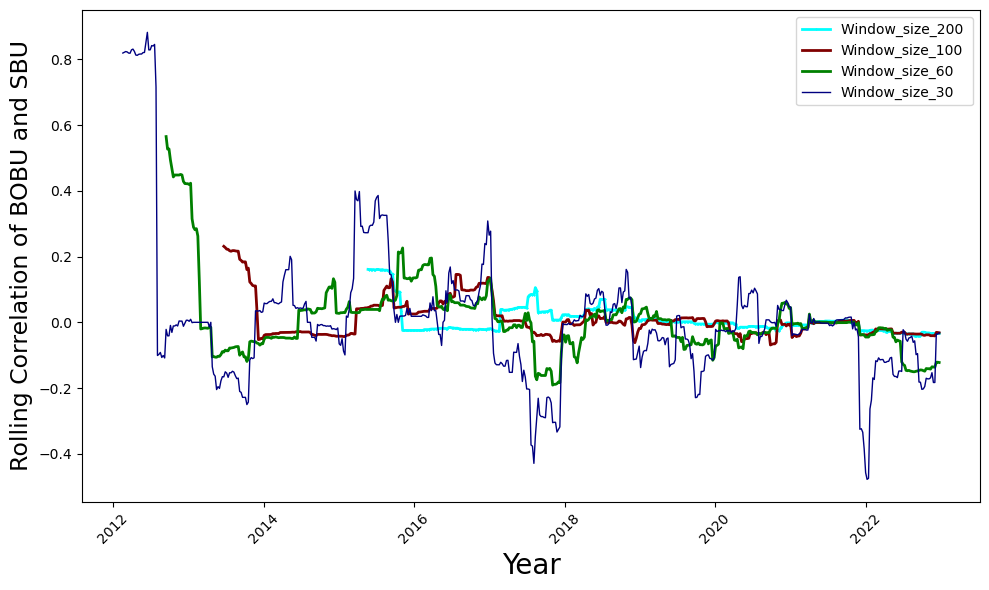}}}\hspace{5pt}
	\hspace{1cm}
	\subfloat[EQTY and KCB.]{%
		\resizebox*{7cm}{!}{\includegraphics{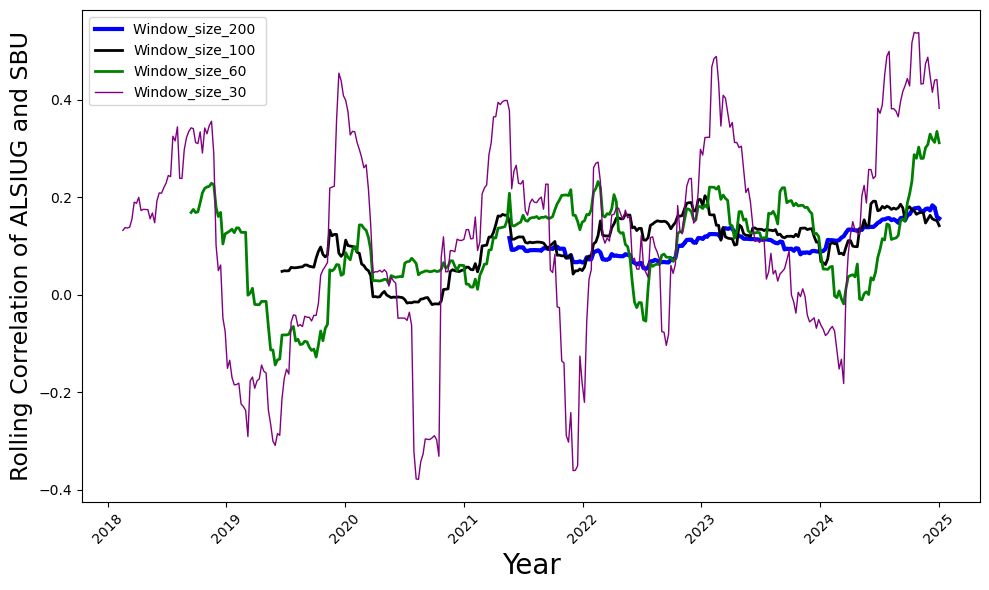}}}\hspace{5pt}
	\caption{Rolling correlation of securities.}
	\label{fig9}
\end{figure}

Figure~\ref{fig8} shows the stock prices of cross listed stocks at USE and their prices have constant prices that are fewer compared to those exhibited by the locally listed companies like SBU, BOBU, DFCU and others. Therefore they possess a high transition probability from state 1 to  state 1.  Equity Group Holdings is listed at both the Nairobi Securities Exchange (NSE) and USE with a ticker symbol of EQTY and EBL respectively. Kenya Commercial bank (KCB) is also listed at both the exchanges. Both these  stocks have a sparse trading activity at USE and a high frequent activity at NSE, they are mainly sold at NSE.
We computed the rolling correlation of  these cross listed stocks using their historical data at NSE  and other securities at USE as shown in  Figure~\ref{fig9} using different window sizes (30,60,100 and 200) using  weekly data. 

SBU and BOBU are  in the same sector and were expected to be highly correlated but it is not the case in our results in Figure~\ref{fig9}, since they are illiquid with constant prices much of the time. They  show a very low correlation, yet they would have been affected by the same economic factors. The correlation of  the stocks is low and negative sometimes.  A window size of 30 was too short to capture the actual correlation of the whole dataset that was used. From the window sizes that are greater than 30, the correlation of the stocks was observed to be very low.

EQTY and KCB  were observed to be highly correlated since their prices are rarely constant as shown in Figure~\ref{fig8}. They also highly correlate with the ALSIUG due to the fact that their probability of staying in the moving state (state 1) is high as well as that of ALSIUG. In addition, their correlation is always positive. This implies that they are not as illiquid as the local listed companies at USE.

\subsubsection{MM-gBm conclusion about correlation }
We simulated two MM-gBm prices  using two correlated gBm and two independent Markov models. Our simulation results show that the pearson's correlation coefficient for the two MM-gBm prices was much lower than that of the correlated gBm prices that were utilized to get the MM-gBm prices, especially for high values of $p,$ the transition probability from state 0 to 0. When $p$ is high, the pearson's correlation coefficient for the two MM-gBm prices, $\rho_{XY}$ is close to that of the correlated gBm prices, $\rho$ that were utilized to generate them.

From the Markov model that we considered with two states; 0 and 1, the steady state probabilities are given by
\begin{eqnarray}
	\pi_0&=&\frac{1-p}{2-p-q},\\
	\pi_1&=&\frac{1-q}{2-p-q}, \\\nonumber 
	\text{and}~~ \pi_0 + \pi_1 &=& 1.
\end{eqnarray}
The measured corelation of a simulation of two MM-gBm prices say; $X$ and $Y$ from two correlated gBm prices  and two independent Markov models was approximated by the following equation;

\begin{equation}\label{eqn:20}
	\rho_{XY} \approx  \frac{(1 - \pi_{0_x})(1 - \pi_{0_y})\rho \sigma_x \sigma_y}{\sqrt{((1 - \pi_{0_x})\sigma_x^2 + (1 - \pi_{0_x})\pi_{0_x} \mu_x^2)((1 - \pi_{0_y})\sigma_y^2 + (1 - \pi_{0_y})\pi_{0_y} \mu_y^2)}},
\end{equation}

where $\sigma_x$ and $\sigma_y$ are the volatility parameters for $X$ and $Y$ respectively, while $\mu_x$ and $\mu_y$ are the parameters for the mean of $X$ and $Y$ respectively.
If the parameters for simulation of the two prices are the same,

$$\rho_{XY} = \frac{\rho \sigma^2 (1-\pi)}{\sigma^2 +  \mu^2},$$ for $\pi=\pi_{0_x}=\pi_{0_y}$, $\mu=\mu_x=\mu_y, \sigma=\sigma_x=\sigma_y$ and when $\mu = 0,$
$$\rho_{XY} = \rho (1-\pi).$$ 

Therefore, we conclude that the low correlation of the stocks at USE is caused by the independent Markov model driving stocks that causes the prices to remain constant much of the time. The lower the number of repetitions in the prices of the stocks, the higher the correlation between them and vice versa.

\section{Conclusions}

The performance of the combination of MM and SDEs (XOU and gBm) was compared with that of gBm and XOU using  daily closing prices of securities at USE.
Our innovative models provide interpretable parameters about trading activity (the MM piece) into which assumptions about the future can be built. (For instance, if traders expect the next week to be quiet,  as would be true in Canada in the week between Christmas and new years eve,  they can change the MM parameters by hand).  The XOU  portions  of the models are  also  highly interpretable as they decompose market dynamics into mean reversion, a mean level, and a volatility term, all of which also bear natural interpretations.  
We tested our models against real data from Ugandan stocks and the main Uganda's index as well.  We had a two, three, four and five parameter model that is;  gBm,  XOU, MM-gBm and MM-XOU respectively.

The MM-gBm outperforms the other models while modelling closing prices of securities at USE especially for a short time period. Our results indicate that adding an MM to an SDE yields a model that is more adherent to the stock prices at USE, as the MM-gBm and MM-XOU on average perform better than gBm and XOU respectively with smaller MAPE values. Additionally, the zero price moves arise from insufficient speculation to accomplish price discovery while  a more liquid market would behave differently. This is due to the low level of speculative trader participation which make it difficult for the market to reflect all available information. Like any data driven models, the MM-gBm and MM-XOU can only describe the future to the extent it behaves like the past. The MAPE values of the MM-gBm were smaller than those of the other models specifically for short time periods that were modelled. However, the MM-gBm and MM-XOU have no relevance in a situation where prices are always moving up and down, this was experimented with ALSIUG and the results from MM-gBm were similar to those of gBm, those form MM-XOU were similar to those of XOU. 
The MM-gBm simulations lead to a conclusion on correlation of the selected illiquid stocks that the lower the number of repetitions in the prices of the stocks, the higher the correlation between them and vice versa. The stocks at USE possess a low correlation due to the constant prices in the prices all the time.
This study can be extended by including more states in the MM rather than 0 and 1 and incorporating other Markov models.

\section*{Appendix}
\section*{Modelling DFCU closing daily prices after COVID 19}

The objective of this appendix is to test the performance of the four models used in this study on DFCU; a USE stock with so many non trading days in a year, and gradual decrease/increase in the prices in a given time period. 
The trading days of DFCU were few compared to those of the other stocks for the data that we used, hence 65 days were used to calibrate the four models after the pandemic.

Data used to calibrate the four models started from $ 5/9/2022 - 3/19/2024 $. In this time period, there was a sudden decrease of the prices of DFCU from 555 to 310 Uganda Shillings as illustrated in Figure~\ref{fig1}. All these 65 days were utilized to calibrate gBm and XOU.  After cleaning the data only 18 days remained 
and used to calibrate the MM-XOU and MM-gBm. DFCU had few trading days in the years that we used, with so many consecutive constant prices compared to the other stocks.  
Some of the MAPE values are summarized in Table~\ref{tab16} and the parameters used were the following;

$$
\begin{array}{lrrrrrr}
	\text{MM-gBm}&:&\mu=-0.0528,&\sigma=0.1639,&p=0.8298,&q=0.5625&\\
	\text{gBm}&:&\mu=-0.0140,&\sigma=0.0859&&&\\
	\text{MM-XOU}&:&\gamma=0.000008,&\sigma=0.1550,&\phi=829.6383,&p=0.8298,&q=0.5625\\
	\text{XOU}&:&\gamma=0.00089,&\sigma=0.0846,&\phi=25.2353
\end{array}
$$

\begin{table}[h]\small \centering
	\caption{MAPE  at different time periods while modelling DFCU stock prices after COVID 19} 
	{
		\begin{tabular}{ccccccc}
			\hline 	\multicolumn{3}{c}{}                                                                                                                               & \multicolumn{4}{c}{\text{MAPE (\%) from different simulations}}\\\hline
			\text{Date}        &\text{Total  days }& \text{Sn}    & \text{MM- gBm }     & \text{gBm}          & \text{MM- XOU}      & \text{XOU}   \\
			\hline
			\multirow{3}{*}{\begin{tabular}[c]{@{}c@{}}3/21/2024\\ -\\ 3/26/2025\end{tabular}}  & \multirow{3}{*}{50}                                   & Sn\_1 & 16.9345      & 27.8461      & 30.5237      & 55.0274     \\ 
			&                                                       & Sn\_2 & 17.1019      & 30.2671      & 21.944       & 53.6829     \\ 
			&                                                       & Sn\_3 & 14.445       & 28.4387      & 30.1475      & 53.7272     \\ \hline
			\multirow{3}{*}{\begin{tabular}[c]{@{}c@{}}3/21/2024\\ -\\ 7/29/2024\end{tabular}} & \multirow{3}{*}{20}                                   & Sn\_1 & 2.1618       & 12.2492      & 7.1281       & 19.2128     \\ 
			&                                                       & Sn\_2 & 4.7748       & 11.411       & 8.121        & 18.0031     \\
			&                                                       & Sn\_3 & 2.3463       & 11.7134      & 6.8203       & 18.5164     \\ 		\hline
	\end{tabular}}
	\label{tab16}
\end{table}

\begin{figure}[h]
	\centering
	\includegraphics[width=11cm]{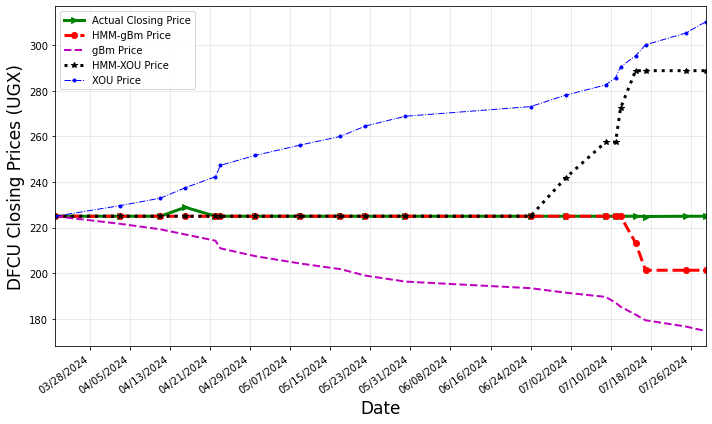} 
	\caption{Modelling stock prices of DFCU after COVID 19  from 3/21/2024 to 7/29/2024  with MAPE values; 0.0883\%, 11.7134\%, 1.0419\% and 18.5164\% for MM-gBm, gBm, MM-XOU and XOU respectively.}
	\label{fig10}
\end{figure}

The  volatility parameter for the non-MM models was smaller than the one for MM-XOU and MM-gBm. For the modelled 50 days from $ 3/21/2024-3/26/2025 $, none of the four models performed well because of a jump in the prices. The MAPE values were too high indicating a diversion of the forecasted values from the actual closing prices of DFCU. However, the models' performance improved when the number of modelled days reduced to 20. The MM-gBm had the lowest MAPE values of all the models followed by MM-XOU model. XOU had the highest MAPE values of all the models. Figure~\ref{fig10} shows the forecasted prices of DFCU and the actual prices from $3/21/2024 -7/29/2024$.

A Kolmogorov-Smirnov test on the distribution of the returns from the forecasted values of the four models  in Figure~\ref{fig10} and the actual returns in the same time period was conducted, results were summarized in Table~\ref{tab17}. The distributions were similar when using MM-gBm and MM-XOU, and they were different for the non-MM models; gBm and XOU.

\begin{table}[H]\small \centering
	\caption{Kolmogorov-Smirnov test on actual and modelled returns of DFCU  after COVID 19 (significance level$=0.01$) }
	{
		\begin{tabular}{ccccc}
			\hline
			& \text{MM-gBm }  & \text{gBm}     & \text{MM-XOU}  & \text{XOU}     \\ \hline
			p-value        & 0.2826 & $1.44e{-13} $ & 0.4879 & $1.53e^{-14}$ \\ 
			Decision & Similar   & Different & Similar  & Different \\ \hline
	\end{tabular}}
	\label{tab17}
\end{table}

\section*{Disclosure statement}
The authors declare no conflict of interest.

\end{document}